\documentclass[twocolumn]{aastex701}

\begin{document}

\title{JADES-GS-z14-1: A Compact, Faint Galaxy at $z\approx14$ with Weak Metal Lines from Extremely Deep JWST MIRI, NIRCam, and NIRSpec Observations}

\correspondingauthor{Zihao Wu}

\author[0000-0002-8876-5248]{Zihao Wu}
\affiliation{Center for Astrophysics $|$ Harvard \& Smithsonian, 60 Garden St., Cambridge MA 02138 USA}
\email[show]{zihao.wu@cfa.harvard.edu}

\author[0000-0002-2929-3121]{Daniel J.\ Eisenstein}
\affiliation{Center for Astrophysics $|$ Harvard \& Smithsonian, 60 Garden St., Cambridge MA 02138 USA}
\email{deisenstein@cfa.harvard.edu}

\author[0000-0002-9280-7594]{Benjamin D.\ Johnson}
\affiliation{Center for Astrophysics $|$ Harvard \& Smithsonian, 60 Garden St., Cambridge MA 02138 USA}
\email{benjamin.johnson@cfa.harvard.edu}

\author[0000-0002-6780-2441]{Peter Jakobsen}
\affiliation{Cosmic Dawn Center (DAWN), Copenhagen, Denmark}
\affiliation{Niels Bohr Institute, University of Copenhagen, Jagtvej 128, DK-2200, Copenhagen, Denmark}
\email{pjakobsen@nbi.ku.dk}

\author[0000-0002-8909-8782]{Stacey Alberts}
\affiliation{Steward Observatory, University of Arizona, 933 N. Cherry Avenue, Tucson, AZ 85721, USA}
\affiliation{AURA for the European Space Agency (ESA), Space Telescope Science Institute, 3700 San Martin Dr., Baltimore, MD 21218, USA}
\email{salberts@arizona.edu, salberts@stsci.edu}

\author[0000-0001-7997-1640]{Santiago Arribas }
\affiliation{Centro de Astrobiolog\'ia (CAB), CSIC–INTA, Cra. de Ajalvir Km.~4, 28850- Torrej\'on de Ardoz, Madrid, Spain}
\email{arribas@cab.inta-csic.es}

\author[0000-0003-0215-1104]{William M.\ Baker}
\affiliation{DARK, Niels Bohr Institute, University of Copenhagen, Jagtvej 155A, DK-2200, Copenhagen, Denmark}
\email{william.baker@nbi.ku.dk}

\author[0000-0002-8651-9879]{Andrew J.\ Bunker }
\affiliation{Department of Physics, University of Oxford, Denys Wilkinson Building, Keble Road, Oxford OX1 3RH, UK}
\email{andy.bunker@physics.ox.ac.uk}

\author[0000-0002-6719-380X]{Stefano Carniani}
\affiliation{Scuola Normale Superiore, Piazza dei Cavalieri 7, I-56126 Pisa, Italy}
\email{stefano.carniani@sns.it}

\author[0000-0003-3458-2275]{St\'ephane Charlot}
\affiliation{Sorbonne Universit\'e, CNRS, UMR 7095, Institut d'Astrophysique de Paris, 98 bis bd Arago, 75014 Paris, France}
\email{charlot@iap.fr}

\author[0000-0002-7636-0534]{Jacopo Chevallard}
\affiliation{Department of Physics, University of Oxford, Denys Wilkinson Building, Keble Road, Oxford OX1 3RH, UK}
\email{chevalla@iap.fr}

\author[0000-0002-2678-2560]{Mirko Curti}
\affiliation{European Southern Observatory, Karl-Schwarzschild-Strasse 2, 85748 Garching, Germany}
\email{mirko.curti@eso.org}

\author[0000-0002-9551-0534]{Emma Curtis-Lake}
\affiliation{Centre for Astrophysics Research, Department of Physics, Astronomy and Mathematics, University of Hertfordshire, Hatfield AL10 9AB, UK}
\email{e.curtis-lake@herts.ac.uk}

\author[0000-0003-2388-8172]{Francesco D'Eugenio}
\affiliation{Kavli Institute for Cosmology, University of Cambridge, Madingley Road, Cambridge, CB3 0HA, UK}
\affiliation{Cavendish Laboratory, University of Cambridge, 19 JJ Thomson Avenue, Cambridge, CB3 0HE, UK}
\email{fd391@cam.ac.uk}

\author[0000-0003-4565-8239]{Kevin Hainline}
\affiliation{Steward Observatory, University of Arizona, 933 N. Cherry Avenue, Tucson, AZ 85721, USA}
\email{kevinhainline@arizona.edu}

\author[0000-0003-4337-6211]{Jakob M.\ Helton}
\affiliation{Steward Observatory, University of Arizona, 933 N. Cherry Avenue, Tucson, AZ 85721, USA}
\email{jakobhelton@arizona.edu}

\author[0000-0003-4512-8705]{Tiger Yu-Yang Hsiao}
\affiliation{Center for Astrophysics $|$ Harvard \& Smithsonian, 60 Garden St., Cambridge MA 02138 USA}
\affiliation{Center for Astrophysical Sciences, Department of Physics and Astronomy, The Johns Hopkins University, 3400 N Charles St. Baltimore, MD 21218, USA}
\affiliation{Space Telescope Science Institute (STScI), 3700 San Martin Drive, Baltimore, MD 21218, USA}
\email{tiger.hsiao@cfa.harvard.edu}

\author[0000-0002-1660-9502]{Xihan Ji}
\affiliation{Kavli Institute for Cosmology, University of Cambridge, Madingley Road, Cambridge, CB3 0HA, UK}
\affiliation{Cavendish Laboratory, University of Cambridge, 19 JJ Thomson Avenue, Cambridge, CB3 0HE, UK}
\email{xj274@cam.ac.uk}

\author[0000-0001-7673-2257]{Zhiyuan Ji}
\affiliation{Steward Observatory, University of Arizona, 933 N. Cherry Avenue, Tucson, AZ 85721, USA}
\email{zhiyuanji@arizona.edu}

\author[0000-0002-3642-2446]{Tobias J.\ Looser}
\affiliation{Center for Astrophysics $|$ Harvard \& Smithsonian, 60 Garden St., Cambridge MA 02138 USA}
\email{tjl54@cam.ac.uk}

\author[0000-0003-2303-6519]{George Rieke}
\affiliation{Steward Observatory, University of Arizona, 933 N. Cherry Avenue, Tucson, AZ 85721, USA}
\email{ghrieke@gmail.com}

\author[0000-0002-5104-8245]{Pierluigi Rinaldi}
\affiliation{Steward Observatory, University of Arizona, 933 N. Cherry Avenue, Tucson, AZ 85721, USA}
\email{prinaldi@arizona.edu}

\author[0000-0002-4271-0364]{Brant Robertson}
\affiliation{Department of Astronomy and Astrophysics University of California, Santa Cruz, 1156 High Street, Santa Cruz CA 96054, USA}
\email{brant@ucsc.edu}

\author[0000-0001-6010-6809]{Jan Scholtz}
\affiliation{Kavli Institute for Cosmology, University of Cambridge, Madingley Road, Cambridge, CB3 0HA, UK} 
\affiliation{Cavendish Laboratory, University of Cambridge, 19 JJ Thomson Avenue, Cambridge, CB3 0HE, UK}
\email{honzascholtz@gmail.com}

\author[0000-0002-4622-6617]{Fengwu Sun}
\affiliation{Center for Astrophysics $|$ Harvard \& Smithsonian, 60 Garden St., Cambridge MA 02138 USA}
\email{fengwu.sun@cfa.harvard.edu}

\author[0000-0002-8224-4505]{Sandro Tacchella}
\affiliation{Kavli Institute for Cosmology, University of Cambridge, Madingley Road, Cambridge, CB3 0HA, UK}
\affiliation{Cavendish Laboratory, University of Cambridge, 19 JJ Thomson Avenue, Cambridge, CB3 0HE, UK}
\email{st578@cam.ac.uk}

\author[0000-0002-9081-2111]{James A.\ A.\ Trussler}
\affiliation{Center for Astrophysics $|$ Harvard \& Smithsonian, 60 Garden St., Cambridge MA 02138 USA}
\email{james.trussler@cfa.harvard.edu}

\author[0000-0003-2919-7495]{Christina C.\ Williams}
\affiliation{NSF National Optical-Infrared Astronomy Research Laboratory, 950 North Cherry Avenue, Tucson, AZ 85719, USA}
\email{christina.williams@noirlab.edu}

\author[0000-0001-9262-9997]{Christopher N.\ A.\ Willmer}
\affiliation{Steward Observatory, University of Arizona, 933 N. Cherry Avenue, Tucson, AZ 85721, USA}
\email{cnaw@as.arizona.edu}

\author[0000-0002-4201-7367]{Chris Willott}
\affiliation{NRC Herzberg, 5071 West Saanich Rd, Victoria, BC V9E 2E7, Canada}
\email{chris.willott@nrc.ca}

\author[0000-0002-7595-121X]{Joris Witstok}
\affiliation{Cosmic Dawn Center (DAWN), Copenhagen, Denmark}
\affiliation{Niels Bohr Institute, University of Copenhagen, Jagtvej 128, DK-2200, Copenhagen, Denmark}
\email{joris.witstok@nbi.ku.dk}

\author[0000-0003-3307-7525]{Yongda Zhu}
\affiliation{Steward Observatory, University of Arizona, 933 N. Cherry Avenue, Tucson, AZ 85721, USA}
\email{yongdaz@arizona.edu}

\begin{abstract}
JWST has shed light on galaxy formation and metal enrichment within 300 Myr of the Big Bang. While luminous galaxies at $z > 10$ often show significant [\ion{O}{3}]$\lambda\lambda$4959,\,5007 emission lines, it remains unclear whether such features are prevalent among fainter, more typical galaxies due to observational limits.  We present deep imaging and spectroscopy of JADES-GS-z14-1 at $z_\mathrm{spec}=13.86^{+0.04}_{-0.05}$, currently the faintest spectroscopically confirmed galaxy at $z\approx 14$. It serendipitously received 70.7 hours of MIRI/F770W imaging in the JWST Advanced Deep Extragalactic Survey (JADES), the deepest MIRI exposure for any high-redshift galaxy to date. Nonetheless, we detect only tentative F770W emission of $7.9\pm2.8$\,nJy at $2.8\,\sigma$ significance, constraining the total equivalent width of [\ion{O}{3}]$\lambda\lambda$4959,\,5007 + H$\beta$ to $520^{+400}_{-380}$ \AA, weaker than most $z > 10$ galaxies with MIRI detections. This source is unresolved across 16 NIRCam bands, implying a physical radius $\lesssim50$\,pc. NIRSpec/PRISM spectroscopy totaling 56 hours reveals no rest-frame ultraviolet emission lines above $3\,\sigma$. Stellar population synthesis suggests a stellar mass $\sim4\times 10^{7}\,\mathrm{M_\odot}$ and a star formation rate $\sim2\,\mathrm{M_\odot\,yr^{-1}}$. The absence of strong metal emission lines despite intense star formation suggests a gas-phase metallicity below 10\% solar and potentially a high escape fraction of ionizing photons. These deep observations provide rare constraints on faint, early galaxies, tracing the onset of chemical enrichment and ionization in the early Universe.
\end{abstract}


\section{Introduction} \label{sec:intro}
JWST has significantly extended the observational frontier, enabling the detection of galaxies within the first 500 Myr after the Big Bang \citep[e.g.,][]{Arrabal2023, Bunker2023GNZ11, Curtis-Lake2023NatAs, Robertson2023NatAs, Wang2023UNCOVER, Carniani_2024, Castellano2024GHZ2, DEugenio2024GSz12, Hainline2024Emission, Harikane2024ApJ, Kokorev2025CAPERS, Naidu2025MoMz14, Napolitano2025AA, Witstok2025GSz13Lya}. These galaxies have revealed unexpectedly high luminosity and prominent nebular emission, suggesting vigorous star formation and rapid metal enrichment in the early Universe \citep{Bunker2023GNZ11, DEugenio2024GSz12, Carniani2025ALMA, Curti2025GSz9, Helton_2024, Schouws2025ALMA}.

Among these high-redshift galaxies, the most luminous objects seem to ubiquitously show significant [\ion{O}{3}]$\lambda\lambda4959,5007$ and H$\beta$ nebular lines, as revealed by JWST/MIRI observations  \citep{Hsiao2024MIRI, zavala2024, Alvarez2025GNz11, Helton_2024}.  Similarly, pre-JWST studies found large equivalent widths (EW) of [\ion{O}{3}] and H$\beta$ in galaxies at $z \approx 7$ (median EW $\approx760\,\mathrm{\AA}$; \citealt{Endsley2021Spitzer}), significantly exceeding those in typical galaxies at $z \approx 2$ (median EW $\approx170\,\mathrm{\AA}$; \citealt{Boyett2022EWlowz}). This phenomenon was initially inferred from Spitzer photometry \citep{Schaerer2009Spitzer, Labbe2013Spitzer, Stark2013Spitzer, Smit2014Spitzer, DeBarros2019Spitzer, Endsley2021Spitzer} and has since been confirmed by JWST spectroscopy \citep[e.g.,][]{Arrabal2023CEERS, Kashino2023EIGER, Sun2023ApJSlitless, Boyett2024EELG, Bunker2024JADES}.

The prominent [\ion{O}{3}] and H$\beta$ lines suggest a high production efficiency of ionizing photons, which may facilitate cosmic reionization in the early Universe provided non-negligible escape fractions \citep[e.g., see discussions in][]{Endsley2021Spitzer}. They also imply that the gas-phase metallicity at $z>10$ has been enriched to $\gtrsim10\%$ solar \citep{Hsiao2024MIRI, zavala2024, Alvarez2025GNz11, Helton_2024}. However, currently known $z>10$ galaxies with MIRI detections are extreme luminosity outliers, exceeding the characteristic luminosity by more than an order of magnitude \citep{Robertson2024ApJ}. To fully understand cosmic reionization and chemical enrichment,  it is essential to observe fainter galaxies, which  dominate the ionizing photon budget and are more representative of the general galaxy population \citep[e.g.,][]{Atek2024reionization, Endsley2024dwarf}.

It remains unclear whether strong nebular lines are also characteristic of less luminous galaxies at high redshift.  Fainter galaxies may have significantly different physical conditions from their luminous counterparts. Observations at lower redshifts, for example, show that low-luminosity galaxies tend to have lower metallicities \citep[e.g.,][]{Curti2023metallicity, Nakajima2023metallicity, Curti2024metallicity, Sarkar2025metallicity} and higher ionizing photon escape fractions \citep{Endsley2023EW, Atek2024reionization, Rinaldi2024Ha, Simmonds2024bursty}, both of which can lead to substantially weaker nebular line emission.

Detecting rest-frame optical emission from faint, high-redshift galaxies with MIRI is extremely difficult due to the long exposure time required. At $z > 10$, the [\ion{O}{3}]$\lambda\lambda4959,5007$ and H$\beta$ nebular lines are redshifted beyond 5$\,\mu$m, making them accessible only through MIRI observations. 
MIRI spectroscopy has so far targeted only galaxies that are relatively bright with exposures of only a few hours \citep{Hsiao2024MIRI, zavala2024, Alvarez2025GNz11}. MIRI imaging has reached fainter objects, such as JADES-GS-z11-0, with $\sim40$ hours of MIRI/F560W integration \citep{Ostlin2025MIDIS, Witstok2025gsz11}. However, galaxies at the redshift frontier are typically even fainter \citep{Robertson2024ApJ}. At $z\approx14$, aside from the overluminous JADES-GS-z14-0 \citep{Carniani_2024}, MoM-z14 and JADES-GS-z14-1 have fluxes of only $20$\,nJy \citep{Naidu2025MoMz14} and $7$\,nJy \citep{Carniani_2024} in NIRCam/F356W, respectively, which are comparable to or fainter than JADES-GS-z11-0. These flux levels underscore the need for deep MIRI observations to explore the earliest and faintest galaxies in the Universe.

In this study, we present ultra-deep MIRI/F770W imaging of JADES-GS-z14-1, a galaxy at $z_\mathrm{spec}=13.86^{+0.04}_{-0.05}$, with its spectroscopic confirmation first reported in \citet{Carniani_2024}. The MIRI imaging has 70.7 hours of on-source integration, which is the deepest MIRI exposure obtained for any high-redshift galaxy to date. JADES-GS-z14-1 is the faintest spectroscopically confirmed galaxy at $z \approx 14$ \citep{Carniani_2024, Naidu2025MoMz14}, with an absolute ultraviolet (UV) magnitude of $M_\mathrm{UV}=-19.0 \pm 0.4$, making it $\sim1$\,mag fainter than the characteristic luminosity $M^*$ of the Schechter function at $z > 12$ \citep[$M^*=-20.2$;][]{Robertson2024ApJ}. Combined with new deep NIRCam and NIRSpec data from the JADES collaboration, we perform a comprehensive analysis of its stellar population and physical conditions.

This paper is structured as follows. In Section~\ref{sec:data}, we describe the observations and data reduction. In Section~\ref{sec:analysis}, we present measurements and analyze the physical properties of the galaxy. Section~\ref{sec:discussion} compares the source with other $z > 10$ galaxies and discuss broad implications. We summarize our findings in Section~\ref{sec:summary}. This study adopts a $\Lambda$CDM cosmology with $H_0 = 68\,\mathrm{ km\, s^{-1} \,Mpc^{-1}}$, $\Omega_m = 0.31$, and $\Omega_\Lambda = 0.69$ according to the final full-mission Planck measurements \citep{Planck2020CMB}.  In this cosmology, 1 arcsec corresponds to 3.36 proper kpc and 50 comoving kpc at $z=13.86$. Throughout this paper, references to the [\ion{O}{3}] emission lines specifically denote the [\ion{O}{3}]$\lambda\lambda4959,5007$ doublet.

\setcounter{footnote}{0}
\section{Observations and Data Reduction}
\label{sec:data}
The observations are primarily conducted as part of JADES and the Observing All phases of StochastIc Star formation (OASIS; PID 5997; PIs: Looser \& D'Eugenio) programs. We refer to \cite{Eisenstein2023JADES, Eisenstein2023Jof} and Looser et al. (in prep.) for details of the observations, and we refer to \cite{Helton_2024, Robertson2024ApJ, Carniani_2024} for descriptions of MIRI, NIRCam, and NIRSpec data reductions\footnote{The data are available at MAST: \dataset[doi: 10.17909/8tdj-8n28]{http://dx.doi.org/10.17909/8tdj-8n28}.}. In the following, we summarize key aspects relevant to this work.

\begin{figure}
    \centering
    \includegraphics[width=0.98\linewidth]{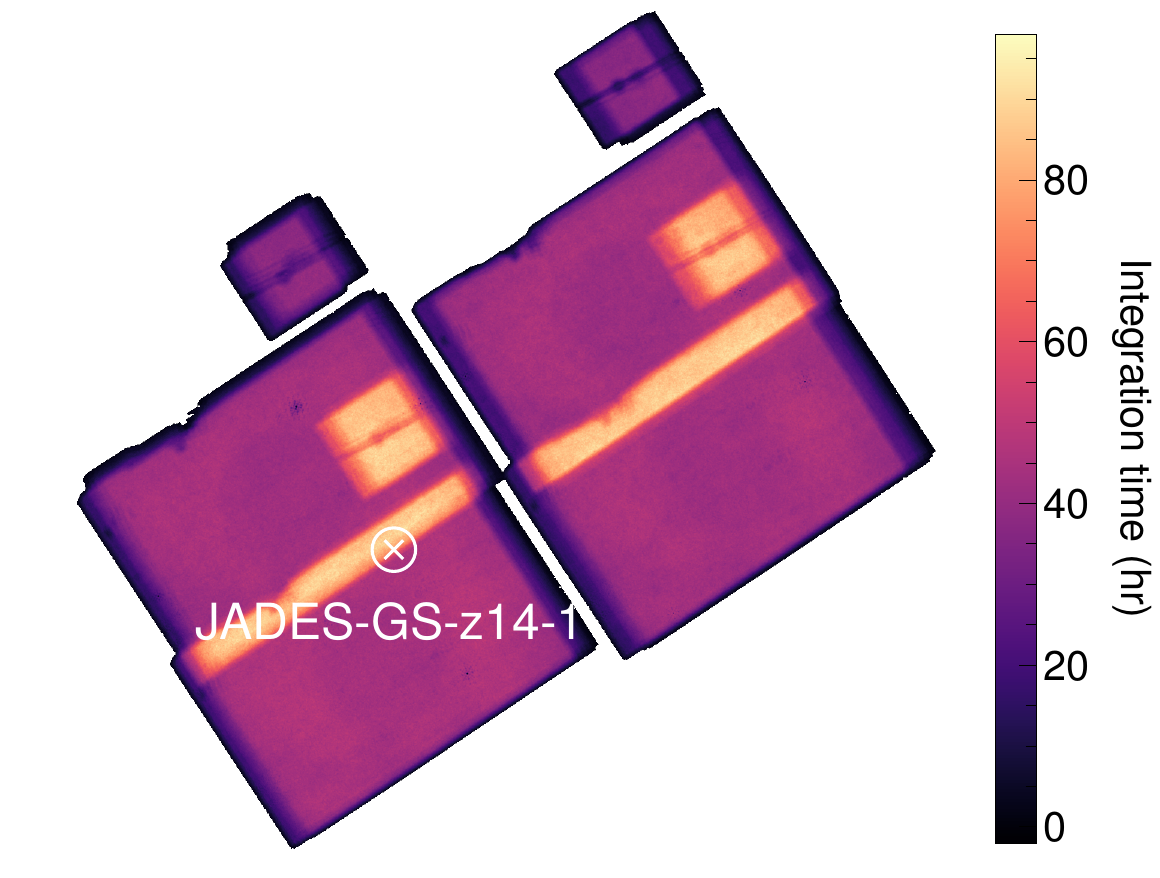}
    \caption{Integration time map of the JADES deep MIRI/F770W parallel field. The cross with circles mark the location of JADES-GS-z14-1, which lies in the overlapping region of two pointings and thus has an integrated on-source exposure time of 70.7 hours.}
    \label{fig:footprint}
\end{figure}

The MIRI/F770W observations are obtained from coordinated parallels to the NIRCam Deep Prime program (PID 1180; PI: Eisenstein), consisting of four pointings with 43.1 hours of open shutter time each. The pointings overlap slightly, and JADES-GS-z14-1 is fortuitously located at the overlap of two pointings, as shown in Figure~\ref{fig:footprint}. Each observation has an integration time of 1361\,s per exposure,  with 114 exposures per pointing using 9 or 4 subpixel dithers. In total, JADES-GS-z14-1 is covered by 187 exposures, resulting in an integrated on-source exposure time of 70.7 hours.

The MIRI data reduction is based on the JWST Calibration Pipeline (v1.16.1) using the Calibration Reference Data System pipeline mapping 1303, following the procedure in \cite{Alberts2024SMILES}, with details in \cite{Helton_2024}. We note that the MIRI noise distribution follows a Gaussian distribution within $3\,\sigma$, but it deviates at higher noise levels. The fraction of $3\,\sigma$ outliers is about three times higher than expected for a Gaussian distribution, based on our analysis of differencing exposure images with the same pointing. We find a persistence artifact in the vicinity of JADES-GS-z14-1 in some exposures, consistent with the behaviors described by \cite{Dicken2024MIRI}, which is caused by a bright star in previous observations. We manually mask the persistence artifact and exclude exposures where the artifact is closer than $0.5''$ to our target.

\begin{figure*}
    \centering
    \includegraphics[width=0.98\linewidth]{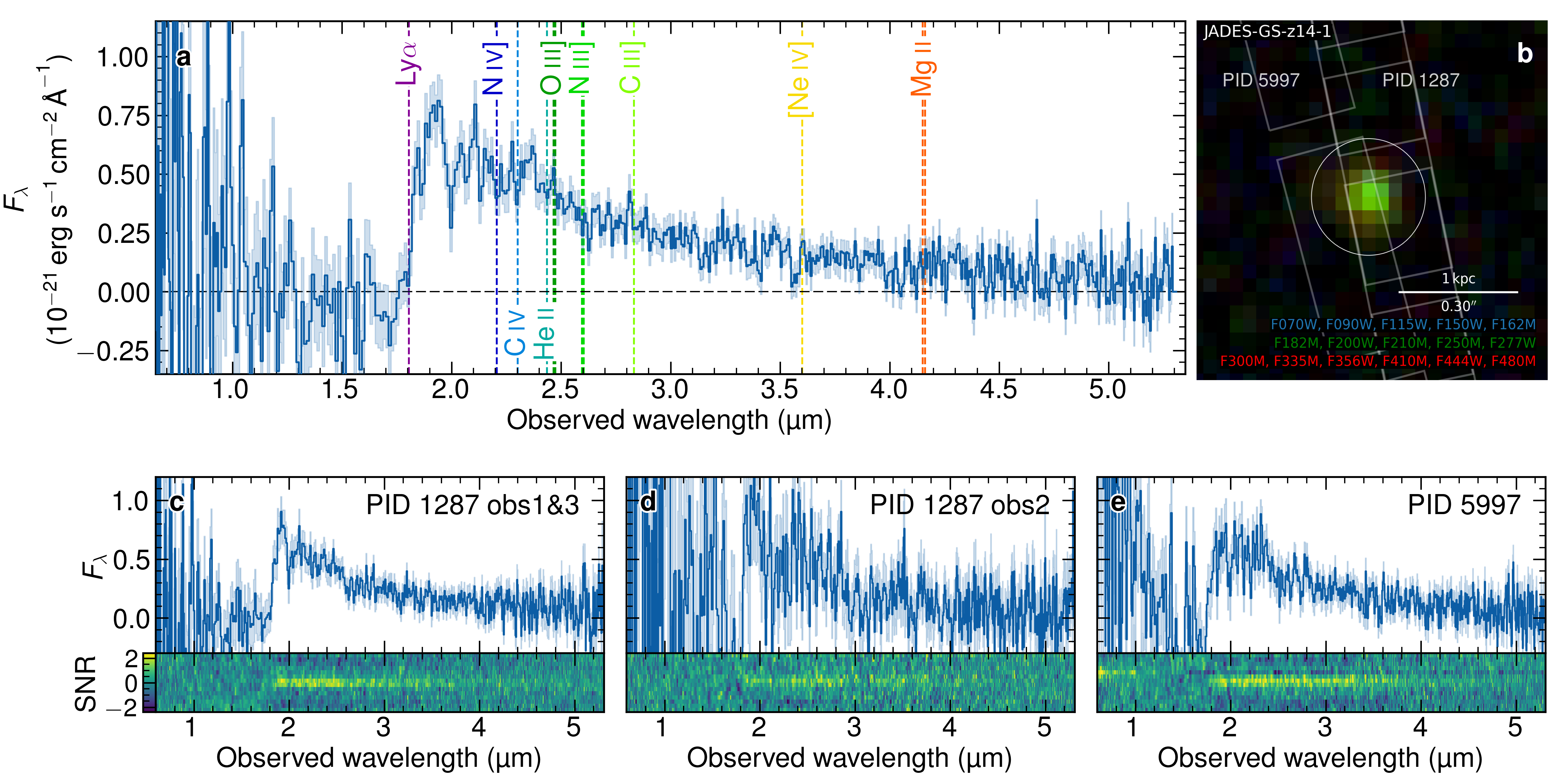}
    \caption{NIRSpec prism spectra of JADES-GS-z14-1, with resolving power of $30\lesssim R\lesssim 300$. (a) Combined 1D spectrum from all observations under PID 1287 and PID 5997. Shaded area in light blue shows the $1\,\sigma$ uncertainty. Vertical dashed lines indicate the expected positions of major emission lines at redshift $z = 13.86$. No emission lines are detected above 3$\,\sigma$ significance. A tentative signal of \ion{O}{3}]$\lambda\lambda1661,1666$ is observed at $2.3\,\sigma$ for $z = 13.80$ and \ion{C}{3}]$\lambda1908$ at $1.9\,\sigma$ for $z = 13.75$. (b) NIRCam false-color image with overlaid slit positions for PIDs 1287 and 5997. (c-e) Individual spectra from (c) PID 1287 first and third dithers, (d) PID 1287 second dither, and (e) PID 5997, after slit-loss correction based on NIRCam photometry. Each panel displays the 1D extracted spectrum at the top and the 2D signal-to-noise ratio (SNR) map at the bottom.}
    \label{fig:spec}
\end{figure*}

The NIRCam data are obtained from JWST Guaranteed Time Observations (GTO) under program IDs 1180 and 4540 (PI: Eisenstein), 1286 (PI: Luetzgendorf), and from General Observer (GO) programs 5997 (PIs: Looser \& D'Eugenio), 2516 (PIs: Hodge \& da Cunha; \citealt{Hodge2025survey}), and 3990 (PI: Morishita; \citealt{Morishita2025survey}). The combined dataset includes eight wide bands and eight medium bands. Details of exposure time and depth in each program are presented in Table~\ref{tab:nircam_observation}. All NIRCam data are reduced using the JADES NIRCam v1.0d pipeline,  which is based on the JWST Calibration Pipeline (v1.14.1) mapping 1228 with improvements including bad pixel masking, wisp and persistence subtraction, and custom sky flats. The long-wavelength (LW) channels of NIRCam have two modules (A and B), which have nearly identical optics and detectors but slightly different throughput; the pipeline therefore processes the LW images in each module separately. The pipeline does not treat the two short-wavelength (SW) modules separately; we therefore average the fluxes across the two detectors for the SW data.

The NIRSpec data are obtained from PID 1287 (PI: Isaak) and PID 5997 (PIs: Looser \& D'Eugenio) using the PRISM/CLEAR grating-filter combination with resolving power of $30 \lesssim R \lesssim 300$ between wavelengths of 0.6\,$\mu$m and 5.4$\,\mu$m. Both observations are conducted in three visits, with each dither producing 24 PRISM subspectra across three separate shutters. For PID 1287, the first and third dithers targeting JADES-GS-z14-1 were conducted on January 10--11, 2024, while the second dither was initially not successfully acquired due to the lock on the guide star being lost, and was repeated on January 12--13, 2025, with a $\sim0.5$-shutter offset perpendicular to the dispersion axis. The three dithers of PID 5997 were all carried out on January 5--7, 2025, with a shifted slit position relative to PID 1287 (see Figure~\ref{fig:spec}b).  The total exposure times are 18.7 hours for dithers 1 and 3 of PID 1287, 9.3 hours for dither 2 of PID 1287, and 28.0 hours for PID 5997, resulting in a combined exposure time three times longer than the observations analyzed by \citet{Carniani_2024}.

The NIRSpec data are reduced using the v4.0 pipeline developed by the ESA NIRSpec Science Operations Team (SOT) and the GTO NIRSpec teams, with advanced background subtraction, rectification, 1D extraction, and spectral combination. A filtering technique is employed to eliminate bad pixels and artifacts. We extract the 1D spectrum using a refined sigma-clipping method applied to three-pixel extractions (0.3$''$) from all available sub-exposures. We apply a wavelength-dependent slit-loss correction based on the source's position within the micro-shutter, assuming a point-source geometry as \cite{Carniani_2024}. We obtain the covariance matrix empirically from bootstrapping of valid wavelength bins of the subspectra, as detailed in Jakobsen et al. (in prep.).

\section{Analysis}
\label{sec:analysis}
We perform a comprehensive analysis of JADES-GS-z14-1 using deep imaging in 16 JWST/NIRCam bands, ultra-deep imaging in MIRI/F770W, and prism spectroscopy from NIRSpec. These data allow us to constrain the photometric and structural properties, search for emission lines, and model the stellar population. In this section, we detail our measurement methods and present the analysis results.

\subsection{NIRCam Photometry}
\label{sec:nircam}

\begin{figure*}
    \centering
    \includegraphics[width=1.0\linewidth]{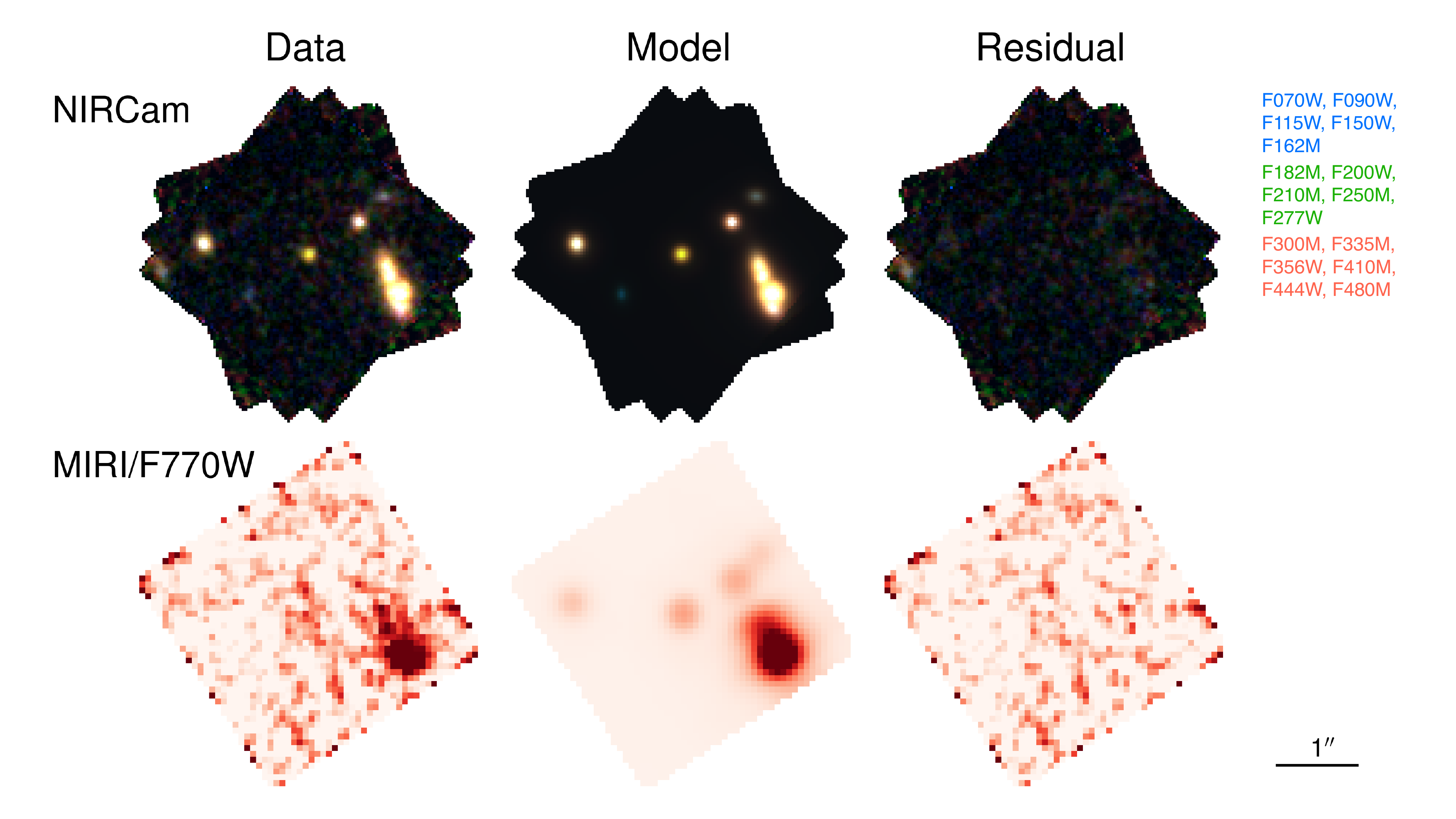}
    \caption{The data, model, and residual for JADES-GS-z14-1 and foreground galaxies across multiple NIRCam bands and the MIRI F770W band. Each panel shows a stack of $2.5''\times 2.5''$ exposure cutouts centered on JADES-GS-z14-1. NIRCam images are shown using false-color stacking with filters assigned to each color channel as annotated on the right. The models are the best-fit S\'ersic profiles convolved with the PSF in each band. Model parameters are inferred from {\tt ForcePho} fitting and are the same across all bands. Images are oriented with North up and East to the left.  A scale bar is shown in the lower right panel.}
    \label{fig:model}
\end{figure*}

We measure the NIRCam fluxes using the package {\tt ForcePho} (B. D. Johnson et al., in prep.; see also \citealt{Robertson2023NatAs, Tacchella2023GNz11, Baker2025ForcePho}), which models the target and nearby sources with S\'ersic profiles and samples model parameters using the Hamiltonian Monte Carlo Markov Chain (MCMC) method. {\tt ForcePho} can properly model barely resolved objects and constrain their sizes \citep{Robertson2023NatAs}. It uses individual exposure images with dithering to achieve subpixel resolution. We perform measurements within $2.5''\times2.5''$ cutout images centered on JADES-GS-z14-1. We fit JADES-GS-z14-1 and nearby objects in the cutouts simultaneously in the 16 NIRCam bands. We directly fit the individual exposure images to avoid correlated noise in the mosaic images, with optimization of local background in each exposure. We adopt {\tt STPSF} (Version 2.0.0; \citealt{Perrin2014WebbPSF}) for the point spread function (PSF) in exposure images,  as empirical PSFs, being derived from dithered mosaics, are not directly applicable to individual exposures. For computational efficiency, {\tt ForcePho} approximates the PSF using a mixture of six Gaussians, a simplification whose impact has been tested and found to be negligible for faint sources \citep{Baker2025ForcePho}. We adopt flat priors for the S\'ersic indices in the range from 0.2 to 8 and for the half-light radii from 1\,mas to 1\,arcsec. We note that a foreground object southwest of JADES-GS-z14-1 comprises four components, and we fit each component with an independent S\'ersic model.  The clean residual image in Figure~\ref{fig:model} shows that the model accurately reproduces the data in all bands.

\begin{deluxetable}{lccccc}
\tablecaption{Photometry of JADES-GS-z14-1 from NIRCam and MIRI/F770W imaging. Model-fitting photometry is performed using the {\tt ForcePho} package for the NIRCam data and a linear least-squares method for the MIRI data, as described in the text. Flux densities from the NIRCam A and B modules are presented separately to account for differences in their transmission curves. We also report NIRCam photometry from 2022, 2023, and 2024 separately, derived using the same model-photometry method. The combined total exposure time across the three years exceeds 6 hours per NIRCam filter; however, exposure times are uneven across years due to data being collected under different observing programs.}
\label{tab:flux}
\tablehead{
\colhead{Filter} & \colhead{Model Fitting} & \colhead{Aperture Photometry} & \colhead{}  & \colhead{Year} & \colhead{}\\ \colhead{} & \colhead{(nJy)} & \colhead{(nJy)} & \colhead{2022} & \colhead{2023} & \colhead{2024}
}
\startdata
F070W   & $-0.4 \pm 0.7$     & $0.6 \pm 1.0$     & \nodata & $-0.4 \pm 0.7$ & \nodata \\
F090W   & $1.0 \pm 0.5$     & $3.3 \pm 0.7$     & $0.8 \pm 0.6$ & $1.3 \pm 0.6$ & \nodata \\
F115W   & $-0.4 \pm 0.5$    & $-0.4 \pm 0.8$    & $-0.8 \pm 0.5$ & $-0.1 \pm 0.6$ & \nodata \\
F150W   & $0.1 \pm 0.4$     & $0.4 \pm 0.6$     & $-0.2 \pm 0.6$ & $0.5 \pm 0.5$ & $0.2 \pm 0.4$ \\ 
F162M   & $-0.6 \pm 0.5$    & $-0.5 \pm 0.8$    & \nodata & \nodata & $-0.6 \pm 0.5$ \\
F182M   & $3.3 \pm 0.4$     & $3.8 \pm 0.8$     & \nodata & \nodata & $3.3 \pm 0.4$ \\
F200W   & $6.4 \pm 0.4$     & $7.5 \pm 0.7$     & $6.6 \pm 0.4$ & $6.3 \pm 0.5$ & $4.8 \pm 1.9$ \\
F210M   & $7.3 \pm 0.5$     & $7.4 \pm 0.9$     & \nodata & \nodata & $7.3 \pm 0.5$ \\\hline
F250MB  & $7.9 \pm 1.0$     & $10.0 \pm 1.1$    & \nodata & $4.9 \pm 3.7$ & $8.2 \pm 1.2$ \\
F277WA  & $10.8 \pm 0.8$    & $10.6 \pm 0.8$    & \nodata & $10.8 \pm 0.8$ & \nodata \\
F277WB  & $7.0 \pm 0.7$     & $7.8 \pm 0.6$     & $7.0 \pm 0.7$ & \nodata & \nodata \\
F300MB  & $8.3 \pm 0.7$     & $9.5 \pm 0.8$     & \nodata & \nodata & $8.3 \pm 0.7$ \\
F335MA  & $6.4 \pm 1.7$     & $7.7 \pm 1.5$     & \nodata & $6.4 \pm 1.7$ & \nodata \\
F335MB  & $6.9 \pm 0.7$     & $7.6 \pm 0.6$     & \nodata & \nodata & $6.9 \pm 0.7$ \\
F356WA  & $7.0 \pm 0.7$     & $7.7 \pm 0.8$     & $5.6 \pm 1.7$ & $7.7 \pm 0.8$ & $9.1 \pm 4.7$ \\
F356WB  & $6.2 \pm 0.6$     & $7.1 \pm 0.6$     & $6.1 \pm 0.7$ & \nodata & $9.8 \pm 3.1$ \\
F410MA  & $5.3 \pm 1.8$     & $4.8 \pm 1.4$     & \nodata & $5.3 \pm 1.8$ & \nodata \\
F410MB  & $5.7 \pm 1.5$     & $5.8 \pm 1.4$     & $5.7 \pm 1.5$ & \nodata & \nodata \\
F444WA  & $6.0 \pm 1.2$     & $5.8 \pm 1.1$     & $1.7 \pm 3.0$ & $7.1 \pm 1.3$ & \nodata \\
F444WB  & $7.2 \pm 1.0$     & $8.0 \pm 1.0$     & $7.1 \pm 1.0$ & \nodata & $16.3 \pm 7.3$ \\
F480MB  & $8.3 \pm 2.8$     & $9.5 \pm 1.9$     & \nodata & \nodata & $8.3 \pm 2.8$ \\\hline
MIRI/F770W   & $7.9 \pm 2.8$     & $8.9 \pm 2.6$     & & & \\\enddata
\end{deluxetable}

\begin{figure*}
    \centering    
    \includegraphics[width=0.9\linewidth]{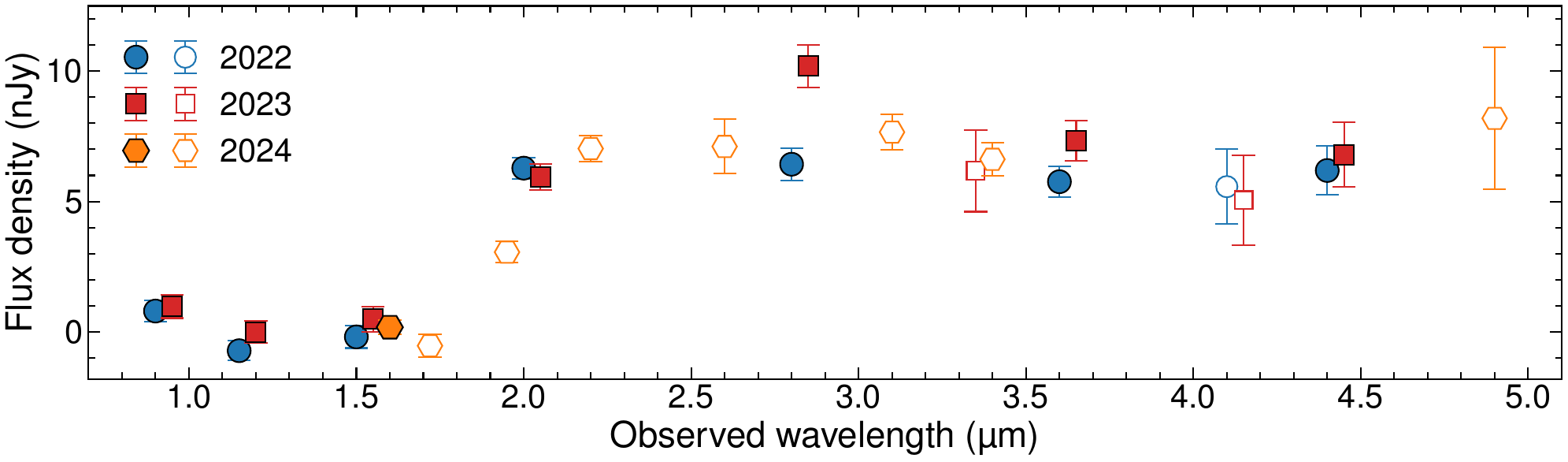}
    \caption{Flux densities of JADES-GS-z14-1 from JWST/NIRCam imaging in 2022, 2023, and 2024, shown across 16 filters spanning observed wavelengths from 0.7 to 4.8 microns. Blue circles, red squares, and orange hexagons represent measurements from 2022, 2023, and 2024, respectively. Wide-band filters are shown with filled markers and medium-band filters with open markers. Flux values from exposures with total integration time below one hour are excluded. For visual clarity, 2023 and 2024 data points are shifted by $+0.05$ and $+0.10$\,$\mu$m, respectively, along the wavelength axis. All observations were conducted between October and December of each year. Flux densities from the NIRCam A and B modules are averaged within each year.}
    \label{fig:variability}
\end{figure*}

We also perform aperture photometry using an aperture of radius $0.1''$, applying aperture correction assuming a point source. The error bar is estimated using random apertures of the same size placed in background regions near the object. Both model fitting and aperture photometry yield consistent results as summarized in Table~\ref{tab:flux}. Notably, the fluxes in the F277WA and F277WB bands, corresponding to two different NIRCam modules, show a 3.8\,nJy ($4\,\sigma$) discrepancy. The two bands have similar but slightly different transmission curves and are observed one year apart. This offset appears in both {\tt ForcePho} and aperture photometry, although the difference shrinks to  2.8\,nJy (3$\,\sigma$) in the latter. It persists across different versions of NIRCam reductions, and is unaffected whether we employ local background fitting. Our inspection of individual exposures reveals no artifacts caused by cosmic rays, scattered light, or unmasked hot pixels. Flat-fielding errors are unlikely to be the cause, as the galaxy was observed at multiple detector locations, averaging out flat-field fluctuations. Furthermore, among 13 nearby sources within $5''$ of JADES-GS-z14-1 with F277W fluxes between 5--20\,nJy, none exhibit a discrepancy between A and B modules that exceeds $2\,\sigma$. We therefore report the discrepancy and conclude that it likely arises from unknown systematics, a rare statistical fluctuation, or extreme variability possibly caused by an active nucleus, supernova, or other transient. A variation of similar sign but smaller amplitude $1.6\pm1.0$\,nJy is observed in the F356W band between the two years, but with only $1.6\,\sigma$ significance. Variability in other bands is no more than 1$\,\sigma$ significance, as shown in Figure~\ref{fig:variability}.

The half-light radius of JADES-GS-z14-1 inferred from the {\tt ForcePho} multiband fitting is less than 10\,mas (84th percentile upper limit). To assess whether the small size reflects detection of only the central brightest region of the galaxy or arises from possible PSF differences, we compare its surface brightness profile with PSF-convolved S\'ersic models of radii of 15\,mas and 30\,mas, as well as with a brown dwarf of similar brightness. Details of this analysis are provided in Appendix~\ref{appendix:profile}.  The comparison of surface brightness profiles constrains the half-light radius to $\lesssim 15$\,mas (50\,pc). Although this constraint is weaker than the {\tt ForcePho} limit---since {\tt ForcePho} achieves higher subpixel resolution by measuring individual dithered exposures---we adopt the more conservative value of 15\,mas, as it is less susceptible to systematics and already sufficient for the scientific interpretations.

\subsection{MIRI Photometry}
\begin{figure*}
    \centering
    \includegraphics[width=1\linewidth]{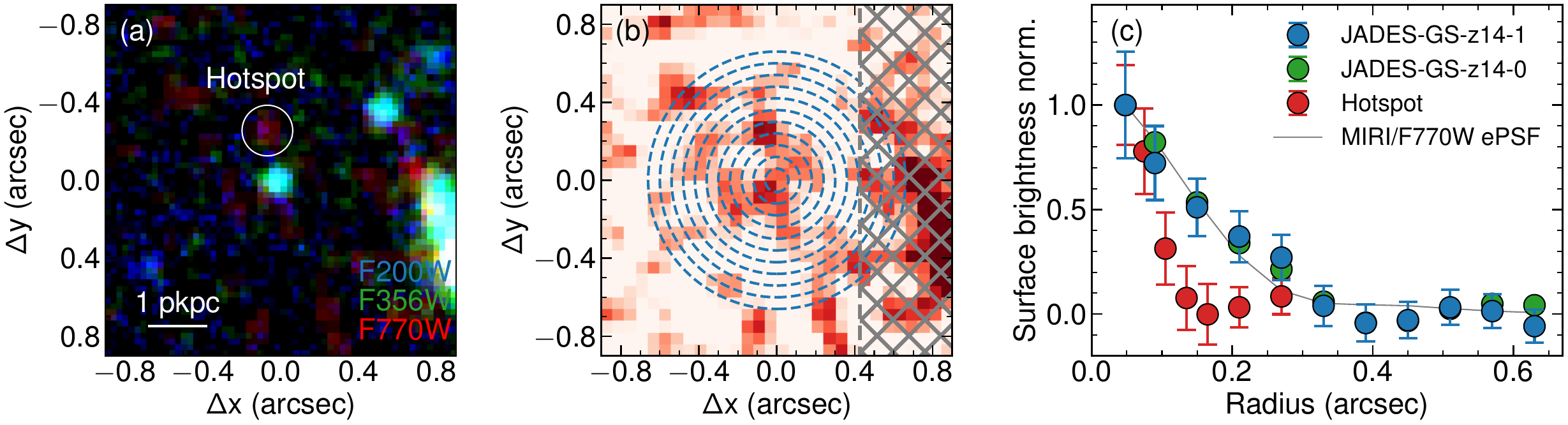}
    \caption{(a) False-color image of JADES-GS-z14-1, showing the relative positions of MIRI and NIRCam emission. The white circle marks a MIRI/F770W hotspot near the target, where the circle has a diameter equal to the FWHM of the MIRI/F770W PSF. (b) JWST MIRI/F770W mosaic image, with blue concentric circles indicating the annuli used for surface brightness profile extraction of JADES-GS-z14-1. Gray hatched regions mark areas with significant contamination by a foreground galaxy. Both the hotspot and the hatched regions are masked during the surface brightness profile measurement of JADES-GS-z14-1. (c) Surface brightness profile of JADES-GS-z14-1 in the MIRI/F770W band in comparison with the MIRI/F770W empirical PSF (ePSF) and the profiles of JADES-GS-z14-0 and the hotspot. The profiles are normalized by the flux in the first aperture. All images are oriented with North up and East to the left.}
    \label{fig:profile}
\end{figure*}

We measure the MIRI/F770W flux using a least-squares method based on the posterior of the NIRCam image fitting, similar to the method applied to JADES-GS-z14-0 \citep{Helton_2024}. We do not fit MIRI and NIRCam images simultaneously with {\tt ForcePho} because this package is only optimized for NIRCam fitting currently. Moreover, the MIRI PSF is not as well understood, and PSF mismatches in MIRI could easily bias galaxy size estimates in a joint fit. Our MIRI photometry adopts the same S\'ersic models as the NIRCam photometry, which ensures model consistency without influencing the inference of structural parameters. Specifically, we forward model all galaxies in MIRI/F770W exposure images, with model parameters, such as coordinates and radii, sampled from the posterior probability of the {\tt ForcePho} fitting in Section~\ref{sec:nircam}.  The S\'ersic models are generated using the {\tt Galsim} package \citep{Rowe2015Galsim} and convolved with an  empirical PSF constructed by combining commissioning data that capture the core and wings of the PSF separately, in order to account for the  cruciform artifacts \citep{Alberts2024SMILES}. We adopt the empirical PSF instead of the {\tt STPSF}  because it captures the cruciform artifacts more accurately, which are prominent effects in the MIRI detector. Direct measurement on individual exposures is critical as the mosaic images exhibit significant pixel covariance and have complex coverage, as our target is located near the edges of some exposures. We determine the fluxes and their uncertainties for JADES-GS-z14-1 and nearby objects using the generalized least-squares formula in Appendix~\ref{appendix}, with simultaneous marginalization of local linear backgrounds in each $2.5''\times2.5''$ exposure image.  Our measurement finds the MIRI/F770W flux of JADES-GS-z14-1 to be $7.9\pm2.8$\,nJy. This result is validated through bootstrapping of the individual exposures and measurements on blank fields, which confirm consistent error estimates. Figure~\ref{fig:model} shows the stacked images, our models, and the residual images.

We also perform aperture photometry with a radius of 0.2$''$, using an annulus with inner and outer radii of 0.4$''$ and 0.5$''$ for background estimation. We measure the fluxes from individual exposures and then combine them with weights given by the inverse of variance. Uncertainties are propagated from the sigma map and background errors. Our aperture photometry yields $8.9\pm2.6\,$nJy, consistent with the model-fitting photometry. The error bar appears slightly smaller because it does not account for possible contamination from neighboring objects. To independently assess the uncertainty, we repeated aperture photometry on randomly selected blank regions with comparable exposure times, located at least 3$''$ away from any source with F444W fluxes above 10\,nJy in the NIRCam catalog. The root mean square (rms) of fluxes in apertures is 2.9\,nJy, in agreement with the error bars in both photometry methods.

We notice a hotspot northeast of our target at a distance of $0.28''$, a separation slightly larger than the FWHM of the MIRI/F770W PSF ($0.269''$). The size of the hotspot is $\sim3$ pixels ($0.18''$), more compact than the PSF FWHM. This hotspot is outside the aperture in aperture photometry, but would significantly increase the flux if a larger aperture were used. Assuming it to be a source, placing a point-source model at the hotspot and jointly fitting it with JADES-GS-z14-1, we find that the flux of the hotspot is $6.3 \pm 2.8$\,nJy, while the flux of our target decreases to $5.9 \pm 2.8$\,nJy. We perform a similar exercise on all NIRCam images with the hotspot location fixed to its MIRI centroid. However, we find no detection above $2\,\sigma$ in any NIRCam band. Given that the hotspot has no counterpart in the deep NIRCam images, appears more compact than the MIRI PSF, and is similar to other hotspots seen in blank regions of the field (Figure~\ref{fig:model} and \ref{fig:profile}b), we suggest that it is a noise outlier due to the non-Gaussian noise distribution discussed in Section~\ref{sec:data}. When masking the hotspot in the model photometry, we find that the flux of JADES-GS-z14-1 decreases by 0.3\,nJy, which is negligible.

Although the hotspot lies outside our photometric aperture, we assess the probability that a similar hotspot could fall within the aperture and dominate the measured flux. To this end, we perform a statistical analysis on the prevalence of hotspots in the nearby field. We perform aperture photometry on random, non-overlapping  blank locations within 2$''$ of JADES-GS-z14-1  and at least 0.5$''$ from sources brighter than $10$\,nJy in any NIRCam band. We find that 13 out of 200 apertures yield fluxes comparable to the hotspot ($\geq6$\,nJy), implying a 6.5\% probability that a similar fluctuation could occur by chance. Therefore, it is not very likely that our photometry is dominated by a random hotspot.

We compare the MIRI/F770W surface brightness profile of JADES-GS-z14-1 with those of the PSF, JADES-GS-z14-0, and the hotspot.  JADES-GS-z14-0 is extended in NIRCam images but barely resolved in MIRI/F770W \citep{Carniani_2024, Helton_2024}. We measure the surface brightness in concentric radial annuli centered on our target with an incremental radius of 0.06$''$.  We mask a 2$\times$2 pixel region near the hotspot and areas located more than 0.36$''$ west of JADES-GS-z14-0 to avoid contamination from the hotspot and extended flux from the foreground galaxies. Figure~\ref{fig:profile}(c) shows that the MIRI/F770W surface brightness profile of JADES-GS-z14-1 is consistent with the PSF and the profile of JADES-GS-z14-0, in contrast to the narrower profile of the hotspot. This distinction supports the fidelity of the MIRI detection.

\subsection{NIRSpec Analysis}
\label{sec:nirspec}

We further correct for slit losses in each NIRSpec pointing using NIRCam photometry, applying a first-order polynomial to account for effects not addressed by the data reduction pipeline. We perform synthetic NIRSpec photometry using {\tt SynPhot} \citep{synphot2020} and compare the results with the NIRCam model photometry. We estimate errors of synthetic photometry by perturbing the spectra according to the noise covariance matrix. After slit-loss correction, we combine the spectra with weights given by their covariance matrices. The combined spectrum and the spectra of the individual pointings are shown in Figure~\ref{fig:spec}.

We model the continuum with a power-law profile, while accounting for intergalactic medium (IGM) and damped Ly$\alpha$ absorption. The power law is fit over the rest-frame wavelength range 1340–2600\,\AA\ using the spectral windows suggested by \citet{Calzetti1994window}. We find a UV slope of $\beta = -2.35 \pm 0.16$, where $\beta$ is defined as the power index in $F_\lambda\propto\lambda^\beta$. The uncertainty of $\beta$ is estimated by randomly perturbing the spectrum according to the covariance matrix.  The IGM absorption and Ly$\alpha$ damping wing are modeled according to \citet{Madau1995ApJ} and \citet{Miralda-Escude98}, with associated parameters set to the best-estimate values obtained from the {\tt Prospector} fit to the spectrum and photometry, as described in Section~\ref{sec:sed}.

We apply a ``redshift sweep" technique \citep{Carniani_2024, Hainline2024Emission} to search for potential emission lines in the continuum-subtracted spectra. This method systematically scans redshifts and evaluates the evidence for line detection, which computes one-sided $p$-values for each candidate line at different redshifts and combines them using Fisher's method to assess the statistical significance of spectroscopic redshifts. However, no emission lines are detected at a $3\,\sigma$ significance level. We find a tentative detection of \ion{O}{3}]$\lambda\lambda1661,1666$ at $2.3\,\sigma$ if $z = 13.80$, and \ion{C}{3}]$\lambda1908$  at $1.9\,\sigma$ if $z = 13.75$. Furthermore, to address potential wavelength calibration offsets among the three pointings, we search emission lines while allowing relative wavelength offsets between individual spectra with a search area of $-0.05$\,$\mu$m to $+0.05$\,$\mu$m in steps of 0.01$\,\mu$m (one channel). Still, no lines are detected above $3\,\sigma$ at any redshift or offset.

We constrain the equivalent widths of emission lines by fitting Gaussian profiles with a fixed FWHM corresponding to the spectral resolution of the NIRSpec prism at the respective line wavelengths, as the lines are unlikely to be spectrally resolved. Line fluxes and uncertainties are calculated using a generalized least-squares approach that makes use of the spectral covariance matrix. To account for redshift uncertainties, we repeat the measurements using redshifts randomly drawn from the posterior distribution of the {\tt Prospector} fitting on the photometry and spectra (Section~\ref{sec:sed}).  As no emission lines are detected, we present the $3\,\sigma$ upper limits of the line flux and rest-frame equivalent width in Table~\ref{tab:emission_limits}. The total flux summed across all listed lines is $(1.6 \pm 1.3) \times 10^{-19}\,\mathrm{erg\,s^{-1}\,cm^{-2}}$, indicating no significant detection.

\begin{deluxetable}{lcc}
\tablecaption{Emission line fluxes and rest-frame equivalent widths measured on the coadded NIRSpec/Prism spectrum. Fluxes are in units of $10^{-19}\,\mathrm{erg}\,\mathrm{s}^{-1}\,\mathrm{cm}^{-2}$, rest-frame equivalent widths ($\mathrm{EW}$) are in \AA. No lines are detected; we therefore report $3\,\sigma$ upper limits.
 \label{tab:emission_limits}}
\tablehead{
\colhead{Emission Line} & 
\colhead{Flux ($10^{-19}$\,erg\,s$^{-1}$\,cm$^{-2}$)} & 
\colhead{EW (\AA)}
}
\startdata
Ly$\alpha$        & $<1.1$ & \\
N\,{\textsc{iv}}]$\lambda1486$ & $<0.7$ & $<10$ \\
C\,{\textsc{iv}}$\lambda1548$  & $<0.8$ & $<11$ \\
He\,{\textsc{ii}}$\lambda1640$ & $<0.7$ & $<12$ \\
O\,{\textsc{iii}}]$\lambda1660$& $<0.6$ & $<11$ \\
N\,{\textsc{iii}}]$\lambda1750$& $<0.6$ & $<11$ \\
C\,{\textsc{iii}}]$\lambda1908$& $<0.5$ & $<13$ \\
Mg\,{\textsc{ii}}$\lambda2795$ & $<0.4$ & $<24$
\enddata
\end{deluxetable}

We detect an absorption feature near $2\,\mu$m at $3.3\,\sigma$ significance, which is also reported in \cite{Carniani_2024}. It occurs at the same wavelength in both the PID 1287 obs1\&3 spectrum (Figure~\ref{fig:spec}c) and the PID 5997 spectrum (Figure~\ref{fig:spec}e), which strengthens its significance. The spectrum in PID 1287 obs2 (Figure~\ref{fig:spec}d) shows tentative evidence, but is too noisy to confirm.  We measure a rest-frame equivalent width of $-13.4 \pm 4.1\,$\AA. If further confirmed, it could be \ion{C}{2}\,$\lambda$1335 doublet absorption with an inflow velocity of $\sim200\,\mathrm{km\,s^{-1}}$. Alternatively, it could arise from foreground intergalactic metal absorption; however, within a 100\,kpc search radius at the corresponding redshift, we find no foreground galaxies with photometric redshifts consistent with producing \ion{Mg}{2} or \ion{Fe}{2} absorption, for galaxies with $M_\mathrm{UV} > -20$. Another possibility is Ly$\alpha$ absorption from inflowing gas, but this would require an extreme velocity of $\sim 2500\,\mathrm{km\,s^{-1}}$, which is usually too high for galaxies \citep{Greif2008first}.

\subsection{Stellar Population Synthesis}
\label{sec:sed}
\begin{figure*}
    \centering
    \includegraphics[width=0.95\linewidth]{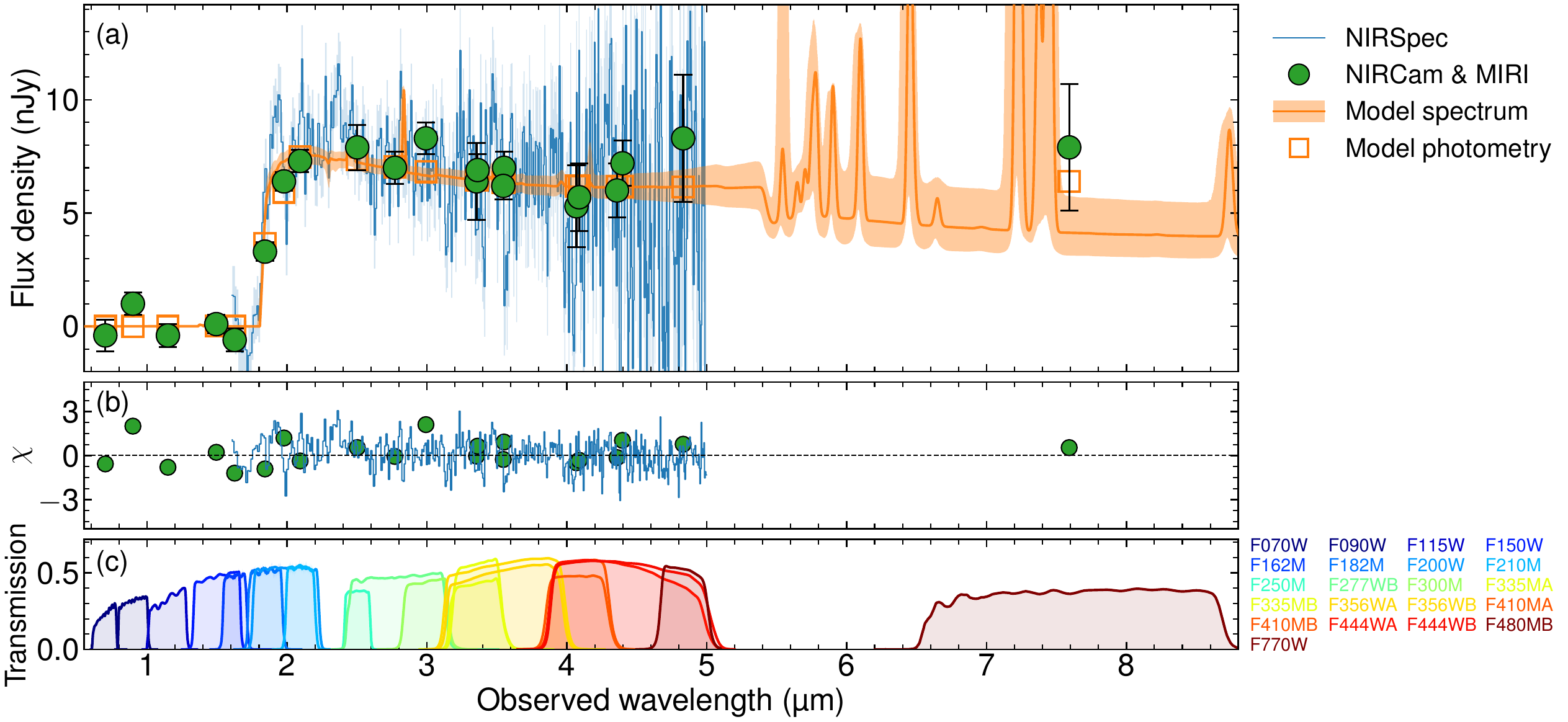}
    \caption{SED modeling of JADES-GS-z14-1 using the {\tt Prospector} package. (a) Observed NIRCam and MIRI/F770W photometry is shown as green filled circles with 1$\,\sigma$ error bars. The coadded NIRSpec spectrum is displayed as a blue step plot, with shaded regions indicating the $1\,\sigma$ uncertainties. The orange curve represents the best-fit spectral flux density, and the shaded orange region denotes the 16th–84th percentile range of the posterior. Orange squares indicate the synthetic photometry corresponding to the best-fit model. (b) Residuals expressed as $\chi$ values, defined as the difference between the observed data and the best-fit model, normalized by the measurement uncertainties. (c) Transmission curves for the relevant filters, with filter names labeled at right. For the NIRCam long-wavelength channel, the photometry and transmission curves for Models A and B are shown separately. The F277WA band is identified as an outlier and is therefore excluded from the modeling and not shown in the plot.}
\label{fig:sed}
\end{figure*}

\begin{figure*}
    \centering
    \includegraphics[width=1\linewidth]{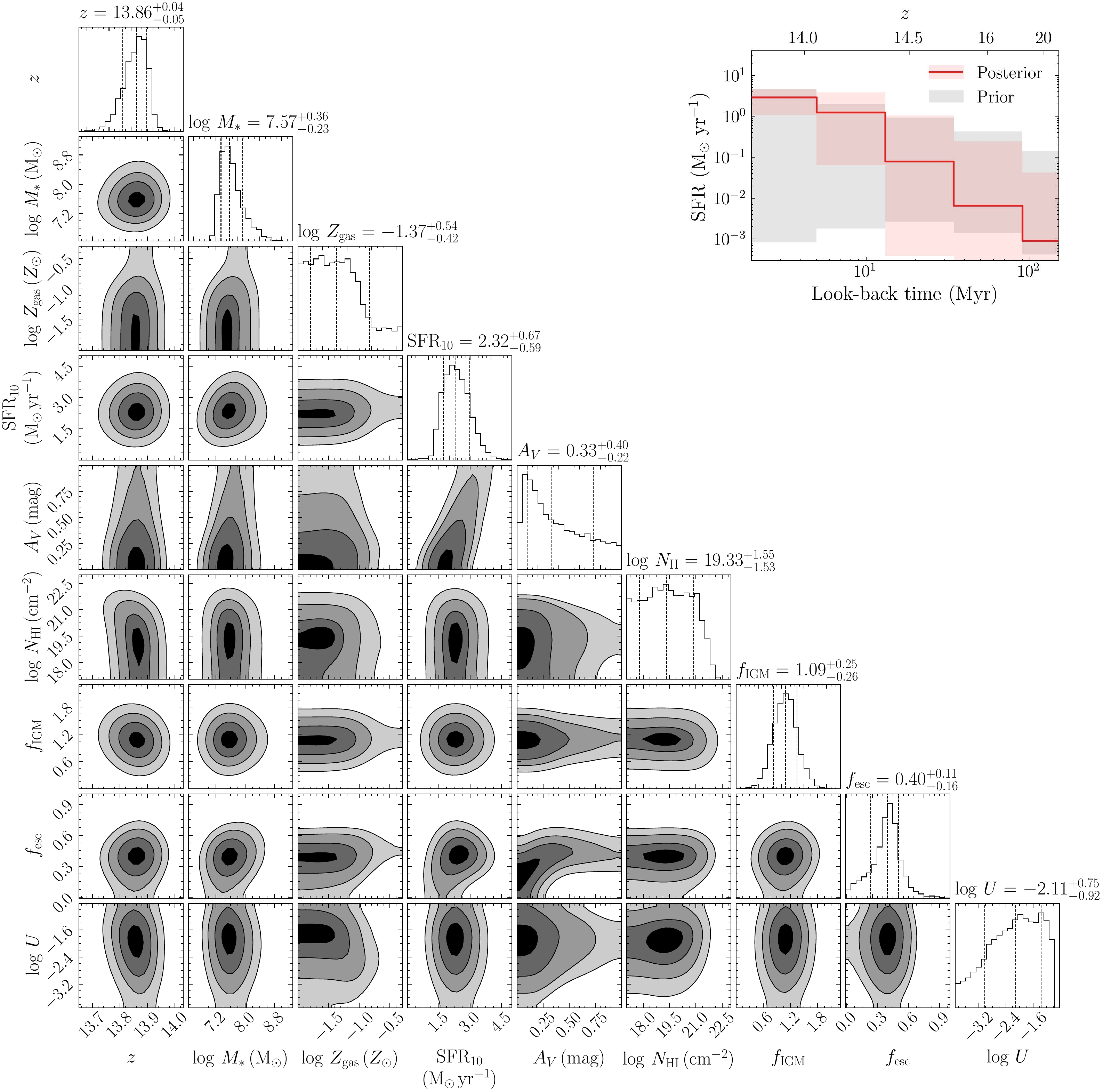}
    \caption{Posterior probability distributions from {\tt Prospector} fitting of NIRSpec, NIRCam, and MIRI data. From left to right, the columns show the posterior distributions for redshift ($z$), stellar mass ($M_*$), gas-phase metallicity ($Z_\mathrm{gas}$), star formation rate in the past 10 Myr (SFR$_{10}$), dust attenuation in the V band ($A_V$) for young stars, the column density of neutral hydrogen ($N_\mathrm{HI}$), the IGM factor ($f_\mathrm{IGM}$), the escape fraction of ionizing photons ($f_\mathrm{esc}$), and the ionization parameter ($U$). The diagonal panels display the 1D marginalized posterior distributions for each parameter, with dashed lines indicating the 16th, 50th, and 84th percentiles. The upper-right panel shows the inferred star formation history (SFH) as a red step plot, with the red shaded region indicating the 16th-84th percentile range of the posterior distribution, and the gray shaded region showing the same percentile range for the prior. The lower x-axis indicates look-back time from the inferred redshift of JADES-GS-z14-1 ($z = 13.86$), while the upper x-axis shows the corresponding redshift. The y-axis represents the star formation rate (SFR).}
    \label{fig:corner}
\end{figure*}

We perform spectral energy distribution (SED) modeling using the Bayesian inference code \texttt{Prospector} \citep{Johnson2021ApJS}, jointly fitting the coadded NIRSpec spectra and the NIRCam and MIRI photometry. The redshift is treated as a free parameter, with IGM absorption and Ly$\alpha$ damping wing simultaneously modeled according to \cite{Madau1995ApJ, Miralda-Escude98, Totani06}.
\texttt{Prospector} employs non-parametric star formation histories (SFHs),  which are flexible for capturing complex star formation conditions at high redshift. 

Stellar population synthesis is performed using Flexible Stellar Population Synthesis \citep[FSPS;][]{Conroy2009ApJ} with Mesa Isochrones and Stellar Tracks (MIST) isochrones \citep{Choi2016ApJ} assuming a \cite{Chabrier2003PASP} initial mass function in the 0.08--120 $\mathrm{M_\odot}$ mass range. We adopt a five-bin non-parametric star formation history (SFH) following \citet{Tacchella2022} and \citet{Turner2025risingSFH}, with the first bin covering a look-back time of 0--5 Myr and the remaining bins logarithmically spaced up to $z=20$. To allow for bursty star formation, which is commonly observed in the early Universe \citep[e.g.,][]{Simmonds2024bursty, Endsley2025Burstiness}, we relax the priors on the star formation ratios between age bins by adopting a Student's t-distribution prior on the $\log\,$SFR ratio, with scale 1 and $\nu = 2$, following \citet{Leja2019}, \cite{Tacchella2022}, and \citet{Carniani2025ALMA}.  To account for the expectation for rising SFHs of high-$z$ galaxies \citep{Turner2025risingSFH}, we adopt a physically-motivated rising SFH prior where the SFR scales as $(1+z)^{-4.5}$, derived from halo mass abundance matching in N-body simulations of Abacus \citep{Maksimova2021ABACUS, Carniani_2024}. Nevertheless, we note that the choice of a rising prior has minimal influence on our results because the prior is very broad and weakly informative. We find that adopting a constant SFH prior changes the inferred SFRs in each bin by less than 5\%.

Dust attenuation is modeled using a two-component dust attenuation model that distinguishes between stars younger and older than 10 Myr. For young stars, we use a power-law attenuation curve with a free slope parameter to model birth-cloud attenuation, adopting a normal prior with a mean of $-1$ and a standard deviation of 0.3 (\citealt{Tacchella2022}, Eq. 4). For older stars, we model the attenuation using the flexible attenuation curves described by \citet{Kriek2013dust}.  The optical depths of the two dust components are independent and assigned flat priors between 0 and 1 \citep[e.g.,][]{Carniani2025ALMA}.

Nebular emission is computed self-consistently following \citet{Byler17}, parameterized by gas-phase metallicity and ionization parameter with flat priors in log space. Stellar metallicity and gas-phase metallicity are allowed to vary independently in our model. We adopt a uniform prior between 0 and 1 for the ionizing photon escape fraction $f_\mathrm{esc}$. In our configuration, a fraction $f_\mathrm{esc}$ of radiation from young stars escapes the galaxy; these escaped photons neither power nebular emission nor experience dust attenuation, while the remaining photons are attenuated, with ionizing photons generating nebular emission. Ly$\alpha$ emission line is excluded in the modeling due to non-detection in the spectra. We include a 2nd-order multiplicative Chebyshev polynomial to match the shapes of the photometry and of the spectrum. Fluxes from the NIRCam A and B modules are treated separately in the SED modeling due to their differences in transmission curves. The F277WA band is excluded from the fit, as it appears to be an outlier inconsistent with both the overall SED shape and the mock photometry derived from the spectrum. Nevertheless, incorporating it has negligible effects on the results. 

Figure~\ref{fig:sed} shows the posterior SED inferred from {\tt Prospector}, along with the MIRI, NIRCam, and NIRSpec data. Overall, the {\tt Prospector} model reproduces the observed fluxes in most bands within $1\,\sigma$ uncertainties. Figure~\ref{fig:corner} presents the posterior distributions for the {\tt Prospector} constraints on the redshift $z$, stellar mass $M_*$,  gas-phase metallicity $Z_\mathrm{gas}$, average star formation rate (SFR) in the past 10 Myr SFR$_{10}$, dust attenuation $A_V$ in the $V$ band, neutral hydrogen column density $N_\mathrm{HI}$, IGM attenuation factor $f_\mathrm{IGM}$, ionizing photon escape fraction $f_\mathrm{esc}$, and ionization parameter $U$. 

The redshift posterior from the {\tt Prospector} fitting yields $z = 13.86^{+0.04}_{-0.05}$, constrained mainly by the Ly$\alpha$ break with IGM absorption and Ly$\alpha$ damping wing modeling. As no emission line is detected, we do not have a redshift estimation informed by emission lines. This error bar is smaller than that in \cite{Carniani_2024}, partly due to the increased NIRSpec depth and the addition of the NIRCam/F182M medium-band observation near the Ly$\alpha$ break. Moreover, \cite{Carniani_2024} incorporates a local ionized bubble and Ly$\alpha$ line emission in the Ly$\alpha$ break modeling, which may change the location of the break and thus increase the redshift uncertainty. The impact of ionizing bubbles on the redshift estimation is relatively small. For instance, a proper line-of-sight span of 0.18$\,$pMpc---the radius of the ionizing bubble for JADES-GS-z13-0-LAE \citep{Witstok2025GSz13Lya}---corresponds to $\Delta z=0.02$ at $z=13.86$, with even less influence on the Ly$\alpha$ break and redshift determination. However, a blended Ly$\alpha$ line may significantly change the apparent break. We find that adding the Ly$\alpha$ line in our model with a free strength broadens the posterior to $z=13.87_{-0.07}^{+0.13}$. Nonetheless, since our data do not show a significant Ly$\alpha$ line and lack the sensitivity to constrain the ionizing bubble, we exclude them from our fiducial SED model to avoid overfitting complex correlated noise near the Ly$\alpha$ break. The redshift of JADES-GS-z14-1 could be underestimated if a weak Ly$\alpha$ line is blended with the continuum.

The estimated stellar mass is $\log\,(M_*/\mathrm{M_\odot}) = 7.57^{+0.37}_{-0.23}$, slightly lower but consistent with the value reported by \citet{Carniani_2024}. This mass is approximately five times smaller than that of JADES-GS-z14-0 \citep{Carniani_2024, Carniani2025ALMA, Helton_2024}, and around three times less massive than MoM-z14 \citep{Naidu2025MoMz14}, consistent with its fainter UV magnitude \citep[$M_{\mathrm{UV}}=-19.0$;][]{Carniani_2024}. Since the observation predominantly traces the rest-frame UV emission, which is dominated by young, massive stars, the stellar mass estimate is sensitive to assumptions about the initial mass function (IMF). If the IMF is more top-heavy, as has suggested to be common in the early Universe \citep{Larson1998IMF, Stacy2016IMF}, the stellar mass would be lower by $\sim0.3$\,dex \citep{Wang2024IMF}.

The inferred gas-phase metallicity is $\lesssim10$\% solar, as indicated by a cliff feature in the posterior probability distribution, above which the probability significantly decreases. This constraint primarily comes from the absence of strong emission lines, indicated by the NIRSpec spectrum and MIRI photometry. Nonetheless, the posterior probability has a tail extending beyond 10\% solar metallicity, with a cumulative probability of 0.16. Such high metallicity requires a high escape fraction (Figure~\ref{fig:corner}), since a high escape fraction implies weaker nebular emission, making the metallicity less constrained.  It is also consistent with the fact that higher metallicities lower the electron temperature, which in turn reduces emission line strengths in our nebular emission model \citep{Byler17}. More stringent constraints on gas-phase metallicity require direct detection of emission lines to determine the electron temperature \citep[e.g.,][]{Hsiao2024MIRI}. The {\tt Prospector} posterior suggests that the stellar metallicity is also  $\lesssim 10$\% solar. However, it is poorly constrained by our photometry and prism spectra due to the lack of resolution to robustly measure metal absorption features \citep[e.g.,][]{Kriek2024metalabsorp}.

The posterior suggests a blue UV slope of $\beta = -2.32\pm 0.08$, consistent with emission from young stars with nebular continuum contribution and modest dust attenuation \citep{Saxena2024UVslope, Topping2024UVslope}. We measure $\beta$ from the ensemble of posterior spectra after {\tt Prospector} fitting, over the rest-frame wavelength range 1340--2600\,\AA \ in the spectral windows defined by \citet{Calzetti1994window}. The posterior value is consistent with the value directly measured from the coadded spectra ($\beta=-2.35\pm 0.16$; Section~\ref{sec:nirspec}), albeit with a smaller error bar due to the addition of photometry data. This slope is slightly redder than the initial value reported in \citet{Carniani_2024}, likely due to the increased data depth and our calibration of slit losses of the NIRSpec spectra using deep NIRCam photometry, which corrects for effects not fully accounted for by the NIRSpec pipeline. The UV slope is bluer than that of JADES-GS-z14-0 ($\beta=-2.20\pm0.07$; \citealt{Carniani_2024}) and is consistent with most galaxies at $z\approx12$ \citep[e.g.,][]{Saxena2024UVslope, Topping2024UVslope}. 

The SED modeling suggests low dust attenuation and a high escape fraction to account for the blue UV slope and weak emission lines. The posterior distribution of dust attenuation ($A_V$) for young stars peaks at 0.1\,mag but extends beyond 0.5\,mag with a declining tail, reflecting its degeneracy with the escape fraction since both parameters shape the UV slope. Moreover, at high escape fractions, most stellar light escapes without attenuation, leaving dust content poorly constrained and largely following the prior. The power-law index of the dust attenuation curve and the optical depth for old stars are not well constrained and thus not shown in Figure~\ref{fig:corner}. The inferred escape fraction $f_\mathrm{esc}=0.40_{-0.17}^{+0.10}$ is consistent with JADES-GS-z13-1-LA ($0.43^{+0.30}_{-0.27}$; \citealt{Witstok2025GSz13Lya}) and larger than that of JADES-GS-z14-0 ($0.11^{+0.09}_{-0.07}$; \citealt{Carniani2025ALMA}). However, if the dust attenuation is $<0.1$\,mag as predicted by simulations \citep[e.g.,][]{Vogelsberger2020TNGdust}, the posterior escape fraction of JADES-GS-z14-1 drops to $0.17^{+0.16}_{-0.10}$.

Our results suggest intense star formation in the past 10 Myr, during which the galaxy has an average SFR $=2.32^{+0.66}_{-0.59}\;\mathrm{M_\odot\,yr^{-1}}$, building up  $67^{+30}_{-40}\%$ of the stellar mass. Given its compact size of $\lesssim 15\,$mas (50\,pc), it implies a star formation surface density of $\Sigma_\mathrm{SFR} \gtrsim 150\;\mathrm{M_\odot\,yr^{-1}\,kpc^{-2}}$, exceeding that of most local starburst galaxies with similar SFRs \citep{Kennicutt2012localSFR}.

The inferred star formation history suggests a rising star formation rate, with a prominent increase around a look-back time of 10 Myr. The logarithmic SFR ratio between adjacent time bins reaches $0.89^{+1.51}_{-0.78}$, suggesting nearly an order-of-magnitude increase in SFR, significantly higher than its prior of $0.09^{+1.31}_{-1.32}$.  The SFH also suggests the population is dominated by young, massive stars formed in the past 10 Myr. This is consistent with the absence of a prominent Balmer break, which typically becomes noticeable only $\sim10$ Myr after starbursts \citep{Trussler2025MNRAS}; such a Balmer break would cause a stronger flux in MIRI/F770W than observed. However, the SFH cannot distinguish between a rising and a declining SFR in the most recent 5 Myr, given the posterior logarithmic SFR ratio of $0.4\pm5.1$ between the most recent two bins.

The posterior spectra suggest mild contributions from [\ion{O}{3}] and H$\beta$ emission lines in the MIRI/F770W flux. From the posterior spectra, we measure a rest-frame equivalent width of $\mathrm{EW}({\mathrm{[O\,III] + H\beta}}) = 520^{+400}_{-380}\,\mathrm{\AA}$, corresponding to a flux excess of $4^{+3}_{-3}$\,nJy in the MIRI/F770W band. This width is slightly smaller but statistically consistent with that of JADES-GS-z14-0 ($670^{+360}_{-130}\,\mathrm{\AA}$; \citealt{Helton_2024}). We note that the inference of equivalent widths depends on the predicted rest-frame optical continuum, which is suppressed in our SED model by strong nebular emission that produces a Balmer jump. If the optical continuum were close to or even higher than the UV level, as expected for a stellar population older than $\sim10$ Myr, the equivalent width would further decrease to $\lesssim 200\,\mathrm{\AA}$.

\begin{figure}
    \centering
    \includegraphics[width=0.95\linewidth]{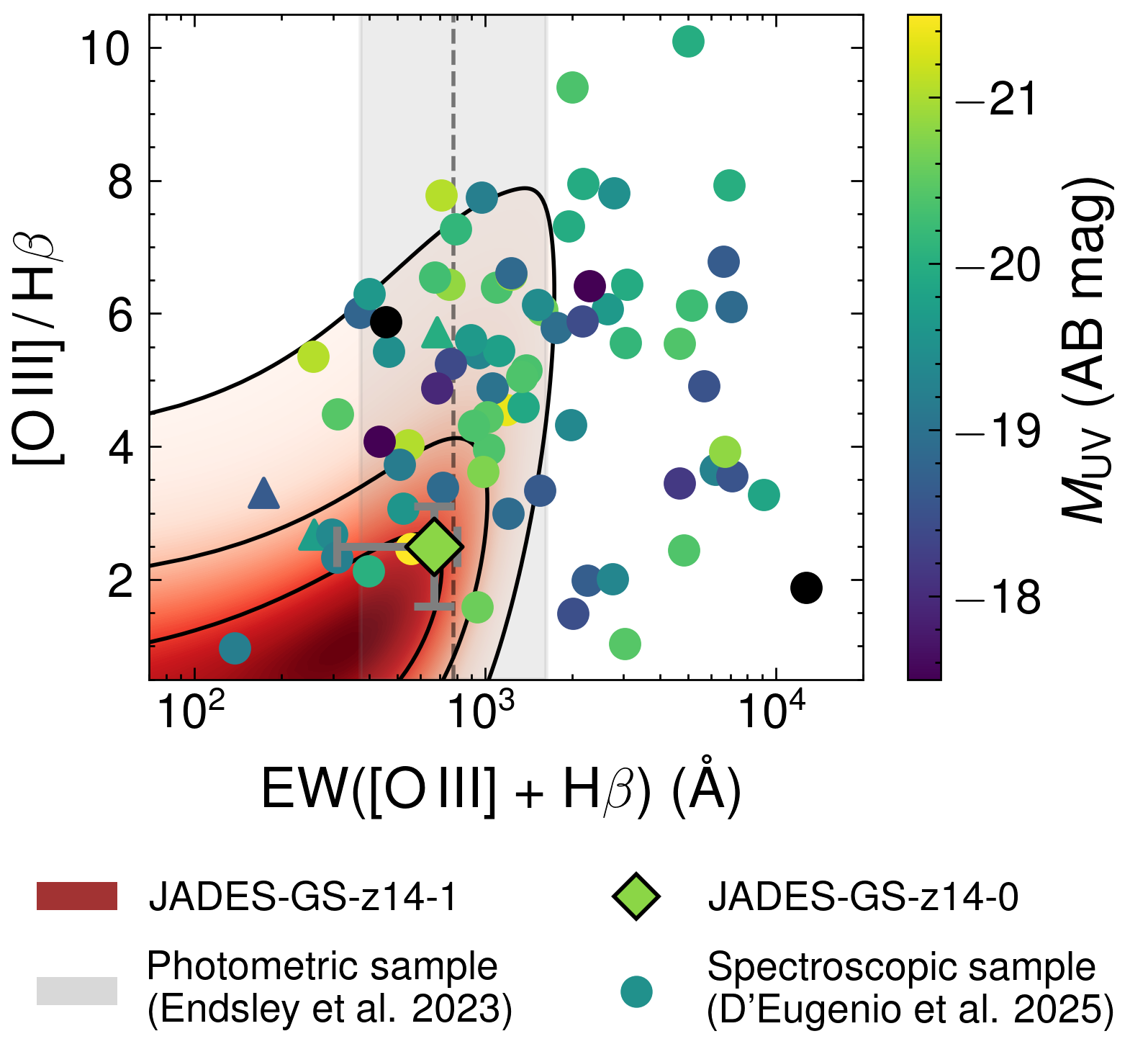}
    \caption{Posterior distribution of the [\ion{O}{3}]/H$\beta$ flux ratio and the equivalent width EW([\ion{O}{3}] + H$\beta$), derived from \texttt{Prospector} fits to MIRI/F770W, NIRCam, and NIRSpec data for JADES-GS-z14-1. The color scale transitioning from red to white indicates the posterior density, with black contours marking the $1\,\sigma$, $2\,\sigma$, and $3\,\sigma$ levels. The black-edged diamond with gray error bars marks JADES-GS-z14-0 from MIRI photometry \citep{Helton_2024}.  The gray shaded region indicates the 16th–84th percentile range of EW([\ion{O}{3}] + H$\beta$) from a photometric $z\approx7$ sample, with the dashed line showing the median \citep{Endsley2023EW}. Colored points represent JADES DR3 galaxies at $z\gtrsim7$ from NIRSpec observations, with color indicating the absolute UV magnitude $M_\mathrm{UV}$; triangles denote $3\,\sigma$ lower limits for galaxies without H$\beta$ detections \citep{DEugenio2025DR3}. The [\ion{O}{3}] lines refer to the [\ion{O}{3}]$\lambda\lambda$4959,5007 doublets.}
    \label{fig:EW}
\end{figure}

We estimate the line ratio [\ion{O}{3}]/H$\beta$ from the {\tt Prospector} posterior spectra. Although not directly constrained by photometry, the line ratio is informed by the combination of a weak total [\ion{O}{3}] + H$\beta$ flux and the expectation of a relatively strong H$\beta$ line, given the high star formation rate and its correlation with H$\beta$ luminosity \citep{DEugenio2025DR3, Helton_2024}. Our measurement on the posterior spectra yields an [\ion{O}{3}]/H$\beta$ ratio of $1.1^{+2.3}_{-0.6}$, suggesting a suppressed [\ion{O}{3}] contribution. As shown in Figure~\ref{fig:EW}, both $\mathrm{EW}$([\ion{O}{3}] + H$\beta)$ and [\ion{O}{3}]/H$\beta$ for JADES-GS-z14-1 are smaller than those of JADES-GS-z14-0, though consistent within $1\,\sigma$. The EW and line ratio for JADES-GS-z14-1 are significantly lower than in most JADES NIRSpec/DR3 galaxies with [\ion{O}{3}] detections, which primarily target  luminous sources \citep{DEugenio2025DR3}, and slightly below the typical values at $z\approx 8$ in photometric samples \citep{Endsley2023EW}. This trend is consistent with the luminosity dependence reported by \cite{Endsley2024dwarf}, where the median EW decreases with UV luminosity. While the EW of JADES-GS-z14-1 is smaller than that of most luminous galaxies, it is consistent with the median for galaxies of similar UV luminosity at $z=7$–9 (median EW([\ion{O}{3}]+H$\beta$) $\approx 580$ \AA\ for $M_\mathrm{UV}\approx-18.6$).

\section{Discussion}
\label{sec:discussion}

\subsection{Comparison with luminous galaxies at  $z>10$}
JADES-GS-z14-1 is only marginally detected in the deep MIRI/F770W image ($7.9 \pm 2.8$\,nJy; $2.8\,\sigma$) and does not show rest-frame UV emission lines in the deep NIRSpec prism spectroscopy. In contrast, the most luminous galaxies at  $z > 10$ exhibit prominent UV metal lines, such as C\,\textsc{iii]} and N\,\textsc{iv]} \citep[e.g.,][]{Bunker2023GNZ11, Castellano2024GHZ2}, and strong MIRI/F770W excesses relative to the NIRCam continuum \citep{Helton_2024}. Some galaxies without MIRI photometry but with spectroscopic [\ion{O}{3}] detections  likely have a comparable excess in the F770W band \citep[e.g.,][]{zavala2024, Alvarez2025GNz11}. To evaluate whether the non-detections in JADES-GS-z14-1 arise from intrinsically weaker nebular lines or from limited sensitivity, we rescale observations of these galaxies as if they were observed at the redshift and brightness of JADES-GS-z14-1.

First, we compare JADES-GS-z14-1 with other $z > 10$ galaxies that show clear MIRI detections by rescaling their SEDs to match the UV brightness of JADES-GS-z14-1 and evaluating their expected F770W fluxes. For JADES-GS-z14-0 \citep{Helton_2024}, we obtain the rescaled F770W flux by downscaling its observed flux according to the ratio of their average NIRCam fluxes between 2--5\,$\mu$m \citep{Carniani_2024}. For GN-z11, GHZ2, and MACS0647-JD, which have MIRI spectroscopic  detections of [\ion{O}{3}] \citep{Hsiao2024MIRI, zavala2024, Alvarez2025GNz11}, we estimate the expected F770W fluxes by adopting the F444W flux \citep{Naidu2022GHZ12, Tacchella2023GNz11, Hsiao2024phot} as the continuum level with a 20\% uncertainty, adding the [\ion{O}{3}] + H$\beta$ line fluxes, and then scaling to the UV brightness of JADES-GS-z14-1. Since none of these objects has significant H$\beta$ detections, we infer the H$\beta$ flux from H$\alpha$ assuming an H$\alpha$/H$\beta$ ratio of 2.86 \citep{Storey1995CaseB}. Except for GHZ2, the available line fluxes include only the [\ion{O}{3}]$\lambda5007$ component. To recover the full doublet strength, we assume a line ratio of [\ion{O}{3}]$\lambda5007/\lambda4959 = 2.98$ \citep{Storey2000line}.

We find that the expected F770W fluxes are $9.5\pm0.7$, $12.1\pm1.5$, $14.3\pm1.6$, and $11.7\pm1.1$\,nJy for JADES-GS-z14-0, GHZ2, GN-z11, and MACS0647-JD, respectively, if they were placed at the redshift of JADES-GS-z14-1 and scaled to match its rest-frame UV luminosity. The error bar accounts for the measurement uncertainty and assumed rest-frame continuum. Given the F770W flux uncertainty of 2.9\,nJy, a more significant detection with SNR = 3.3--5.1  would be expected if emission lines in JADES-GS-z14-1 were comparable to those of the reference galaxies. Therefore, the lack of detection suggests an intrinsically lower equivalent width of [\ion{O}{3}] and H$\beta$ lines.  Alternatively, a reduced continuum level by a stronger Balmer jump could also lower the F770W flux of JADES-GS-z14-1.  However, a stronger Balmer jump is typically associated with a larger EW([\ion{O}{3}] $+$ H$\beta$), and the combined effect would usually increase---not decrease---the F770W flux \citep{Katz2024nebular}. Therefore, a weaker nebular line strength is a more plausible explanation for the faintness in F770W.

Second, we simulate how the spectrum of GN-z11 would appear if observed at the brightness, redshift, and noise level of JADES-GS-z14-1. GN-z11 is one of the most well-studied galaxies at $z>10$ with clearly detected rest-frame UV emission lines such as N~\textsc{iv}], N~\textsc{iii}], C~\textsc{iii}], and Mg~\textsc{ii}, and a high-SNR continuum spectrum \citep{Bunker2023GNZ11}. Although the two galaxies may differ in many physical properties, they are both compact, and their nearly identical UV slopes provide a fair comparison in line equivalent width measurement.  We redshift the GN-z11 spectrum from \citet{Bunker2023GNZ11} to $z = 13.86$, matching the redshift of JADES-GS-z14-1, and rescale its flux to match the observed brightness. As the nebular emission lines in GN-z11 are unresolved, redshifting alters their apparent widths in the prism spectra. To account for this, we first remove the original emission lines, including \ion{N}{4}], \ion{N}{3}], and \ion{C}{3}], and re-inject them using Gaussian profiles with widths matched to the spectral resolution at the corresponding observed wavelengths. The injected line EWs are kept consistent with the original GN-z11 measurements. We then add noise consistent with the JADES-GS-z14-1 NIRSpec data by sampling from its covariance matrix, generating 100 mock spectra that incorporate different noise realizations. Each mock spectrum is processed using the same line-fitting and measurement pipeline as applied to JADES-GS-z14-1.

The resulting signal-to-noise ratios for the injected features indicate that \ion{C}{3}] would be detected with equivalent widths of $12.5 \pm  3.4\,$\AA \ (3.6\,$\sigma$), \ion{N}{4}] at $6.6 \pm 3.0\,$\AA \ (2.2$\,\sigma$), and \ion{N}{3}] with $6.7 \pm 3.1\,$\AA \ $(2.2\,\sigma$), while all other lines fall below $1.5\,\sigma$. While these correspond to marginal detections, the absence of \ion{C}{3}] in JADES-GS-z14-1 suggests its \ion{C}{3}] emission is weaker than that of GN-z11. The \ion{C}{3}] line is widely observed in high-redshift star-forming galaxies, and extrapolations from $z = 5$--11 samples predict an EW(\ion{C}{3}]) $\sim20$\,\AA\ at $z = 14$ \citep{Roberts-Borsani2024CIII}. In contrast, our observations constrain JADES-GS-z14-1 to EW(\ion{C}{3}]) $<13$\,\AA\ at $3\,\sigma$, suggesting that its \ion{C}{3}] emission is mildly weaker than expected for a typical star-forming galaxy at this epoch.

\subsection{Implications of weak metal lines}
The absence of strong [\ion{O}{3}] + H$\beta$ emission in JADES-GS-z14-1 stands in contrast to the intense nebular lines often observed in star-forming galaxies at $z\approx8$, where rest-frame optical equivalent widths often exceed 700\,\AA \ \citep{Endsley2021Spitzer, Tacchella2023EW, Bunker2024JADES, Begley2025EW, DEugenio2025DR3}. The strong emission line is commonly attributed to bursty star formation enabled by the short dynamical timescale relative to the stellar feedback timescale in the early Universe \citep{Tacchella2016MNRAS, Faucher2018timescale, Orr2019timescale, McClymont2025arXiv}.  In JADES-GS-z14-1, however, the EW([\ion{O}{3}] + H$\beta$) $= 520^{+400}_{-380}\,\mathrm{\AA}$ is lower than the majority of galaxies observed at $z\approx8$, despite indications of similar bursty star formation. Even more striking is its unusually low  [\ion{O}{3}]/H$\beta$, since the H$\beta$ emission is expected to be high given the high star formation rate. These results suggest that the physical conditions of JADES-GS-z14-1 may be different from typical galaxies at the Epoch of Reionization.

A natural explanation is that a very low gas-phase metallicity suppresses [\ion{O}{3}] emission. Our {\tt Prospector} modeling indeed suggests a median gas-phase metallicity of merely $5\%\,Z_\odot$, which is consistent with the stellar mass-metallicity relation at $z\approx10$ \citep{Curti2023metallicity, Kashino2023EIGER, Curti2024metallicity, Sarkar2025metallicity}.  Although the interstellar medium may be $\alpha$-enhanced \citep[e.g.,][]{Arellano2022chemical}, the absolute oxygen abundance could still be too low to generate prominent [\ion{O}{3}] lines. This is consistent with expectations at $z \sim 14$, when the Universe was only 300--400 Myr old, and galaxies are expected to have undergone few enrichment events. Cosmological simulations predict widespread metallicities of $Z \lesssim 0.1\,Z_\odot$ in $z > 5$ galaxies, especially in low-mass galaxies ($M_* \lesssim 10^8\,\mathrm{M_\odot}$) that retain largely pristine gas reservoirs \citep[e.g.,][]{Wise2012metal, Torrey2019metal}. Observationally, many galaxies at $z > 6$ have confirmed metallicities below 5\% solar \citep[e.g.,][]{Schaerer2022metal, Curti2023metallicity, Nakajima2023metallicity, Curti2024metallicity, Laseter2024metal, Cullen2025metal, Li2025metal, Tang2025arXiv, Willott2025metal}, and in some cases even $< 1\%\,Z_\odot$ \citep[e.g.,][]{Nakajima2023metallicity, Vanzella2023metal, Hsiao2025metal, Morishita2025lowz}. If anything, such metal-poor systems should become more common at earlier times.

It is also possible that the weakness is due to a high escape fraction of ionizing photons \citep[e.g.,][]{Endsley2024dwarf, Hainline2024Emission, Baker2025miniquench}. The extremely compact morphology of JADES-GS-z14-1 could facilitate efficient escape of ionizing photons, reducing the reprocessing of ionizing radiation into nebular lines.  Additionally, the low stellar mass of JADES-GS-z14-1  implies a low-mass dark matter halo with a shallow potential well. Such halos are more susceptible to gas loss from stellar feedback, which can deplete the interstellar medium and suppress nebular line emission. If these conditions are representative of faint galaxies at $z > 12$, it would imply that such systems contribute to cosmic reionization through efficient leakage of ionizing photons into the intergalactic medium, perhaps creating small, early ionized bubbles.

Another possibility is that some of the most massive O-type stars have already died, reducing the supply of ionizing photons. Our SFH inference cannot distinguish between a rising and a declining SFR within the past 5 Myr. Since massive O-type stars live only a few Myr, some may have already ended their lives, leaving the UV light dominated by slightly less massive stars that produce fewer ionizing photons \citep{Endsley2024dwarf}. Nonetheless, the SFH indicates that this galaxy has likely not been quenched for a long period, unlike the mini-quenched galaxies observed at $z \approx 8$---which show extremely weak or absent rest-frame UV line emission \citep[e.g.,][]{Looser2024Natur, Baker2025miniquench, Looser2025miniquench}---because the faint MIRI flux suggests no significant Balmer break has formed. An old stellar population is unlikely also because older post-burst galaxies are intrinsically faint \citep{Endsley2025Burstiness}; for instance, the mini-quenched galaxy in \citet{Looser2024Natur} would appear at only $\sim3$\,nJy if placed at $z=14$, falling below detection limits, and would likely be even fainter after accounting for galaxy mass evolution.

Finally, we note that our estimate of the [\ion{O}{3}] + H$\beta$ equivalent width relies on MIRI broadband photometry that blends line and continuum emission. Uncertainties in the continuum shape, especially in the presence of nebular continuum, affect the emission line inference. Direct spectroscopic measurements, such as MIRI Low Resolution Spectroscopy (LRS) observations scheduled for JADES-GS-z14-0 (PID 8544; PI: Helton), would allow for precise line flux measurements and stringent constraints on the gas-phase metallicity and ionization conditions. However, the extreme faintness of JADES-GS-z14-1 would require a prohibitively long exposure time of  $\sim600$ hours for MIRI/LRS to achieve a 3\,$\sigma$ detection of [\ion{O}{3}] emission lines. A more feasible path forward is to identify gravitationally lensed galaxies at similar redshifts with MIRI spectroscopy. Promising examples of this strategy have already emerged \citep[e.g.,][]{Hsiao2024MIRI}, and such objects will be critical to investigating faint galaxies at cosmic dawn.

\subsection{Implications of extremely compact size}
JADES-GS-z14-1 is extremely compact, with a half-light radius $\lesssim15$\,mas (50\,pc), which is remarkably small even among galaxies at $z>10$ \citep[see summary in][]{Harikane2025clump}, and smaller than predicted by cosmological simulations of early galaxy formation \citep[e.g.,][]{Liu2017size, Shen2024size, McClymont2025size}. Among the other galaxies at $z\approx14$, JADES-GS-z14-0 has a half-light radius of 280\,pc \citep{Carniani_2024}, and MoM-z14 has a radius of 74\,pc \citep{Naidu2025MoMz14}, both larger than that of JADES-GS-z14-1. While most galaxies at $z>10$ follow a size evolution consistent with extrapolations from lower redshifts \citep{Shibuya2015size, Harikane2025clump}, JADES-GS-z14-1 falls well below this trend, with a size more than an order of magnitude smaller than typical expectations. \citet{Harikane2025clump} suggest that compact galaxies at $z>10$ may have a different evolutionary pathway compared with extended galaxies, as the former usually show stronger emission lines. However, the weak emission lines observed in JADES-GS-z14-1 complicate this interpretation and may indicate greater diversity among compact galaxies.

In the mass-size relation, JADES-GS-z14-1 is over an order of magnitude smaller than  galaxies in the local Universe of similar stellar masses \citep{Lange2015masssize}. Even compact, starbursting galaxies in the local Universe, such as Green Pea galaxies \citep{Cardamone2009Greenpea}, rarely reach sizes below 300\,pc \citep{Kim2021greenpea}. Intriguingly, the mass and size of JADES-GS-z14-1 resemble those of ultra-compact dwarf galaxies (UCDs; \citealt{Phillipps2001UCD}). For example, M60-UCD1 has a stellar mass of $\sim 2 \times 10^8\,\mathrm{M_\odot}$ and an effective radius of 14\,pc \citep{Strader2013UCD}, broadly consistent with those inferred for JADES-GS-z14-1. \cite{Strader2013UCD} find that the stellar population of M60-UCD1 is older than 10 Gyr and thus could be a remnant of early galaxies.  However, it is unlikely to be a direct descendant of JADES-GS-z14-1, as stars in M60-UCD1 have nearly solar metallicity, while the metallicity of JADES-GS-z14-1 is $<10\%$ solar.

JADES-GS-z14-1 is consistent with stellar clumps observed in high-redshift galaxies in the mass–size relation (e.g., \citealt{Cava2018clump, Kalita2024clump}). This raises the possibility that star formation in JADES-GS-z14-1 may experience similar conditions to those in stellar clumps. Many observations have shown that star formation in the early Universe may concentrate in compact, clump-like structures (e.g., \citealt{Fujimoto2024grape, Harikane2025clump}). Such clump-driven modes of star formation can increase star formation efficiencies and may help explain the high observed star formation rates in $z > 10$ galaxies.

The compact size of JADES-GS-z14-1 implies a very short dynamical timescale of $\sim1$ Myr at the effective radius, which would be even shorter if the galaxy contains a substantial gas fraction. This timescale is more than three times shorter than that of JADES-GS-z14-0 and nearly two orders of magnitude shorter than those of typical local galaxies \citep{Lange2015masssize}. Such a brief timescale suggests that star formation could proceed in rapid, intense bursts, especially when the dynamical timescale is shorter than the lifetimes of massive O-type stars ($\sim3$--10 Myr), which typically end in supernovae that regulate star formation through feedback. In this regime, gas can collapse rapidly, triggering a burst of star formation before supernova feedback can interrupt this process \citep{Tacchella2016MNRAS, Faucher2018timescale, Orr2019timescale, McClymont2025arXiv}. Indeed, the star formation rate of JADES-GS-z14-1 implies that nearly 70\% of the stellar mass has formed within the past 10 Myr, consistent with a recent starburst.

The extreme compactness of JADES-GS-z14-1 may also have implications beyond star formation. Its high stellar density could increase the frequency of dynamical interactions, potentially facilitating the formation of an intermediate-mass black hole via runaway stellar collisions and early black hole mergers \citep{Greene2020ARA&A}. The compact structure may also be linked to active galactic nuclei activity, which will be explored in the following section. Furthermore, its compact morphology, luminosity, and the absence of UV emission lines are broadly consistent with predictions from the dark star hypothesis, which suggests that early stars may be powered by dark matter annihilation, and will eventually collapse and form massive black hole seeds \citep{Spolyar2009DStar, Freese2016Dstar}.

\subsection{Possibility of an active galactic nucleus (AGN)}
Although JADES-GS-z14-1 is too faint to permit the detection of definitive AGN spectral features, several observational characteristics remain consistent with the AGN scenario. First, the source is unresolved in deep NIRCam imaging. Second, the rest-frame UV slope of $\beta=-2.32\pm0.08$ is in agreement with the theoretical slope of a standard accretion disk ($F_\lambda\propto\lambda^{-7/3}$; \citealt{Shakura1973AGNpowerlaw}). This raises the possibility that the UV continuum may be dominated by AGN emission.

Due to the faintness of the source, we cannot detect or meaningfully constrain UV emission lines characteristic of AGNs, such as [Ne\,\textsc{iv}]~$\lambda2423$ and C\,\textsc{ii}$^*$~$\lambda1335$, nor do we detect significant C\,\textsc{iv}~$\lambda1549$ absorption indicative of AGN-driven outflows. The tentative absorption feature at 2\,$\mu$m (rest-frame 1346\,\AA) is not likely caused by C\,\textsc{iv} because it implies an unrealistic outflow velocity of $\sim 0.1c$. A direct search for blueshifted C\,\textsc{iv} absorption within $\pm2000$\,km\,s$^{-1}$ yields a maximum EW of $-5.1 \pm 3.4$\,\AA \ at an outflow velocity of 1300\,km\,s$^{-1}$, comparable to GN-z11 with EW $\approx-5$\,\AA, which \cite{Maiolino2024GNz11} suggest to host an AGN based on the detection of [Ne\,\textsc{iv}] and C\,\textsc{ii}$^*$ lines. However, our signal is not statistically significant and suffers from the look-elsewhere effect since we search the signal in a large parameter space.  Overall, no UV spectral feature confirms or strongly favors AGN activity, but none of the available constraints conclusively rule it out either.

Assuming, hypothetically, that the observed UV luminosity is powered entirely by AGN activity, we can estimate the black hole mass given its UV absolute magnitude $M_\mathrm{UV}=-19\,$mag \citep{Carniani_2024}, provided a bolometric correction factor of 5 \citep{Netzer2019AGNbolo}. Under Eddington-limited accretion, this corresponds to a black hole mass of $\sim 1.4 \times 10^6\,\mathrm{M_\odot}$, implying a seed mass of $\sim 6 \times 10^4\,\mathrm{M_\odot}$ at $z=20$. This seed mass is consistent with direct collapse scenarios \citep{Begelman2006BH}, but significantly larger than that expected from stellar remnants or runaway collisions in dense star clusters \citep{Greene2020ARA&A}. If instead the black hole is accreting at a super-Eddington rate of $\lambda_\mathrm{Edd} \approx 5.5$---similar to GN-z11 \citep{Maiolino2024GNz11}---the black hole mass drops to $\sim 2.5 \times 10^5\,\mathrm{M_\odot}$. In comparison, GN-z11 is slightly more massive with an inferred black hole mass of $1.6 \times 10^6\,\mathrm{M_\odot}$ accreting at $\sim 5.5$ times the Eddington rate \citep{Maiolino2024GNz11}.

In the AGN scenario, an interesting question is whether JADES-GS-z14-1 could represent an earlier evolutionary stage of systems like GN-z11. However, simple forward extrapolation of GN-z11 black hole growth from $z = 14$ to $z = 11$ suggests that its progenitor would have had a black hole mass of $<10^2\,\mathrm{M_\odot}$ at $z = 14$, far below the inferred mass of JADES-GS-z14-1 under the AGN hypothesis. Even assuming Eddington-limited growth, the GN-z11 progenitor would only reach $\sim 10^5\,\mathrm{M_\odot}$ at $z = 14$, still below the $\sim 1.4 \times 10^6\,\mathrm{M_\odot}$ inferred for JADES-GS-z14-1 at Eddington accretion. Therefore, unless more than 90\% of the UV luminosity of JADES-GS-z14-1 originates from stellar emission, or unless its accretion rate changes over time, it is unlikely to be a direct progenitor of GN-z11. Conversely, if JADES-GS-z14-1 is indeed solely powered by AGN activity and maintains a constant accretion rate, it would evolve into a black hole $>10^7\,\mathrm{M_\odot}$ by $z = 11$ under any assumption of accretion rates, which is more massive than the black hole mass of GN-z11 suggested by \cite{Maiolino2024GNz11}.

\section{Summary}
\label{sec:summary}
We present extremely deep JWST observations of JADES-GS-z14-1, the faintest spectroscopically confirmed galaxy at redshift $z\approx14$ to date. This object received an unprecedented 70.7-hour MIRI/F770W integration, 16-band NIRCam photometry, and 56 hours of NIRSpec/PRISM spectroscopy as part of the JADES program. Despite intense star formation activity, we detect only tentative F770W emission ($7.9\pm2.8$\,nJy at 2.8$\,\sigma$) and no significant rest-frame UV emission lines from the spectra. The galaxy appears spatially unresolved in all NIRCam bands, indicating a half-light radius  of $\lesssim 15$\,mas (50\,pc). The increased data depth allows us to constrain the redshift of JADES-GS-z14-1 to $z = 13.86^{+0.04}_{-0.05}$ by fitting the Ly$\alpha$ break while accounting for IGM absorption and Ly$\alpha$ damping wings.

Joint modeling of NIRCam, MIRI, and NIRSpec data using \texttt{Prospector} yields a low stellar mass of $\log\,(M_*/\mathrm{M_\odot}) = 7.57^{+0.37}_{-0.23}$ and gas-phase metallicity of $\log\,(Z_{\mathrm{gas}}/Z_\odot) = -1.37^{+0.55}_{-0.42}$. The inferred star formation history shows a steep rise in the last 10 Myr, during which the galaxy has an average star formation rate of $ 2.3_{-0.6}^{+0.7}\,\mathrm{M_\odot\,yr^{-1}}$ and forms nearly 70\% of its stellar mass. The galaxy shows a UV continuum slope of $\beta = -2.32 \pm 0.08$, consistent with moderate nebular continuum or dust attenuation. The dust attenuation posterior peaks at $A_V=0.1$\,mag but extends beyond 0.3\,mag due to degeneracy with the escape fraction of ionizing photons. The escape fraction is inferred to be high, around 40\%, but could decrease to $\sim15\%$ if the galaxy is dust-free. Even this lower value implies a significant contribution to ionizing photon escape, potentially marking the early formation of ionized bubbles in the intergalactic medium.

As a faint, low-mass galaxy, JADES-GS-z14-1 lacks the extreme nebular line emission commonly observed in the most luminous $z>10$ galaxies. The absence of strong [\ion{O}{3}] and H$\beta$ emission---evidenced by a [\ion{O}{3}] + H$\beta$ equivalent width of $520^{+400}_{-380}$\,\AA---stands in contrast to the intense nebular lines typically seen in luminous galaxies at $z > 10$. When we scale down the SEDs of these brighter galaxies to match the UV brightness of JADES-GS-z14-1, the \ion{C}{3}]$\lambda1908$ line and the MIRI/F770W flux are expected to be detected at $\gtrsim 3.5\,\sigma$ significance. The lack of such detections in both spectroscopy and MIRI imaging for JADES-GS-z14-1, therefore, suggests intrinsically weaker nebular metal lines. Our SED modeling interprets the weakness of the metal lines as evidence of a low gas-phase metallicity of $\sim5$\% solar that suppresses the nebular line emission.

JADES-GS-z14-1 is more compact than most known $z>10$ galaxies and at least an order of magnitude smaller than the predictions based on the size evolution at $z\approx8$ or the local mass–size relation. Its mass and size are more consistent with ultra-compact dwarf galaxies and stellar clumps. This compactness may enhance star formation efficiency, facilitating its recent starburst activity. It also raises the possibility of AGN activity. The UV slope is consistent with that of a standard accretion disk, and if the UV emission is dominated by an AGN radiating at the Eddington limit, the black hole mass could be $\sim10^6\,\mathrm{M_\odot}$. However, because of the faintness of the object, we find no compelling evidence for AGN activity.

These extremely deep JWST observations offer a rare, detailed look at a faint, low-mass galaxy at the redshift frontier. The weakness of metal emission lines suggests that strong nebular lines are not ubiquitous in early galaxies. These results underscore the diversity of early galaxy properties and provide key constraints on the metal enrichment, star formation, and ionizing conditions during the first 300 Myr after the Big Bang.

\software{
ForcePho (B. D. Johnson et al., in prep.),
FSPS \citep{Conroy2009ApJ},
GALFIT \citep{Peng2002GALFIT, Peng2010GALFIT},
GalSim \citep{Rowe2015Galsim},
MIST isochrones \citep{Choi2016ApJ},
Nautilus \citep{Lange2023NAUTILUS},
Photutils \citep{Bradley2025},
Prospector \citep{Johnson2021ApJS},
STPSF \citep{Perrin2014WebbPSF}.
}

\begin{acknowledgments}
We are grateful to the anonymous referee for helpful comments. We thank Xiaojing Lin, Yifei Jin, Lisa Kewley,  Peixin Zhu, Ana Sofia Uzsoy, Seiji Fujimoto, Daniel Stark, and Takahiro Morishita for insightful conversations. This work is based on observations made with the NASA/ESA/CSA James Webb Space Telescope. The data were obtained from the Mikulski Archive for Space Telescopes at the Space Telescope Science Institute, which is operated by the Association of Universities for Research in Astronomy, Inc., under NASA contract NAS 5-03127 for JWST. These observations are associated with programs 1180, 1286, 1287, 2516, 3990, 4540, and 5997. The authors acknowledge the
Morishita team (PID 3990) for developing their observing program with a zero-exclusive-access period.

DJE, BDJ, KH, JH, ZJ, BR, GR, FS, CNAW, and YZ acknowledge support from the NIRCam Science Team contract to the University of Arizona, NAS5-02015. DJE is also supported as a Simons Investigator and by NASA through a grant from the Space Telescope Science Institute, which is operated by the Association of Universities for Research in Astronomy, Inc., under NASA contract NAS5-03127. The Cosmic Dawn Center (DAWN) is funded by the Danish National Research Foundation under grant DNRF140. SA acknowledges support from the JWST Mid-Infrared Instrument (MIRI) Science Team Lead, grant 80NSSC18K0555, from NASA Goddard Space Flight Center to the University of Arizona. WMB gratefully acknowledges support from DARK via the DARK fellowship and research grant (VIL54489) from VILLUM FONDEN. SAr acknowledges grant PID2021-127718NB-I00 funded by the Spanish Ministry of Science and Innovation/State Agency of Research (MICIN/AEI/ 10.13039/501100011033). AJB and JC acknowledge funding from the “FirstGalaxies” Advanced Grant from the European Research Council (ERC) under the European Union’s Horizon 2020 research and innovation programme (Grant agreement No. 789056). SC acknowledges support by European Union’s HE ERC Starting Grant No. 101040227 – WINGS. ECL acknowledges support of an STFC Webb Fellowship (ST/W001438/1). FDE acknowledges support by the Science and Technology Facilities Council (STFC), by the ERC through Advanced Grant 695671 “QUENCH,” and by the UKRI Frontier Research grant RISEandFALL.  TJL gratefully acknowledges support from the Swiss National Science Foundation through an SNSF Mobility Fellowship and from the NASA/JWST Program OASIS (PID 5997). JS acknowledges support by the Science and Technology Facilities Council (STFC), ERC Advanced Grant 695671 ``QUENCH". ST acknowledges support by the Royal Society Research Grant G125142. JAAT acknowledges support from the Simons Foundation and JWST program 3215. Support for program 3215 was provided by NASA through a grant from the Space Telescope Science Institute, which is operated by the Association of Universities for Research in Astronomy, Inc., under NASA contract NAS 5-03127. The research of CCW is supported by NOIRLab, which is managed by the Association of Universities for Research in Astronomy (AURA) under a cooperative agreement with the National Science Foundation.

This research made use of the lux supercomputer at UC Santa Cruz which is funded by NSF MRI grant AST 1828315. This work was carried out using resources provided by the Cambridge Service for Data Driven Discovery (CSD3) operated by the University of Cambridge Research Computing Service (www.csd3.cam.ac.uk), provided by Dell EMC and Intel using Tier-2 funding from the Engineering and Physical Sciences Research Council (capital grant EP/T022159/1), and DiRAC funding from the Science and Technology Facilities Council (www.dirac.ac.uk).
\end{acknowledgments}

\appendix
\restartappendixnumbering 
\twocolumngrid
{
\section{NIRCam Observation Information}
Here we summarize the NIRCam observations used in this work, including exposure times, depths, and program IDs for each filter. Table \ref{tab:nircam_observation} lists these details.

\begin{deluxetable*}{lcccccccc}
\caption{NIRCam observations used in this work, including exposure times, depths, and program IDs for each filter. }
\tablehead{
\colhead{Filter} & \colhead{PID 1180} & \colhead{PID 1286} & \colhead{PID 2516} & \colhead{PID 3990} & \colhead{PID 4540} & \colhead{PID 5997} & \colhead{Total Exp.} & \colhead{5$\,\sigma$ Depth} \\
\colhead{} & \colhead{(hour)} & \colhead{(hour)} & \colhead{(hour)} & \colhead{(hour)} & \colhead{(hour)} & \colhead{(hour)} & \colhead{(hour)} & \colhead{(nJy)}
}
\label{tab:nircam_observation}
\startdata
F070W    & \nodata & 1.57 & \nodata & \nodata & 5.23 & \nodata & 6.80 & 4.41 \\
F090W    & 3.15    & 2.36 & \nodata & \nodata & \nodata & \nodata & 5.51 & 4.06 \\
F115W    & 3.87    & 3.15 & \nodata & \nodata & \nodata & \nodata & 7.02 & 3.36 \\
F150W    & 3.15    & 2.36 & \nodata & \nodata & 0.09 & 6.87 & 12.47 & 2.08 \\
F162M    & \nodata & \nodata & \nodata & \nodata & \nodata & 6.87 & 6.87 & 4.22 \\
F182M    & \nodata & \nodata & \nodata & \nodata & \nodata & 6.87 & 6.87 & 3.50 \\
F200W    & 3.15    & 2.36 & 1.05 & \nodata & 0.26 & \nodata & 6.82 & 2.58 \\
F210M    & \nodata & \nodata & \nodata & \nodata & \nodata & 6.87 & 6.87 & 3.90 \\
F250MB   & \nodata & \nodata & \nodata & 0.93 & \nodata & 6.87 & 7.80 & 2.20 \\
F277WA   & \nodata & 2.36 & \nodata & \nodata & \nodata & \nodata & 2.36 & 2.02 \\
F277WB   & 3.15    & \nodata & \nodata & \nodata & \nodata & \nodata & 3.15 & 1.69 \\
F300MB   & \nodata & \nodata & \nodata & \nodata & \nodata & 6.87 & 6.87 & 1.69 \\
F335MA   & \nodata & 1.57 & \nodata & \nodata & \nodata & \nodata & 1.57 & 3.49 \\
F335MB   & \nodata & \nodata & \nodata & \nodata & \nodata & 6.87 & 6.87 & 1.58 \\
F356WA   & \nodata & 2.36 & 0.52 & \nodata & 0.09 & \nodata & 2.97 & 1.77 \\
F356WB   & 3.15    & \nodata & \nodata & \nodata & 0.17 & \nodata & 3.32 & 1.69 \\
F410MA   & \nodata & 2.36 & \nodata & \nodata & \nodata & \nodata & 2.36 & 3.97 \\
F410MB   & 3.15    & \nodata & \nodata & \nodata & \nodata & \nodata & 3.15 & 3.45 \\
F444WA   & \nodata & 3.15 & 0.52 & \nodata & \nodata & \nodata & 3.67 & 2.79 \\
F444WB   & 3.87    & \nodata & \nodata & \nodata & 0.09 & \nodata & 3.96 & 2.73 \\
F480MB   & \nodata & \nodata & \nodata & \nodata & \nodata & 6.87 & 6.87 & 5.69\\
\enddata
\tablecomments{5$\,\sigma$ depths are measured from random apertures on nearby blank sky, using a 1$''$ radius aperture and applying point-source aperture corrections. The quoted depths are from the coadded images combining all programs.}
\end{deluxetable*}

\section{NIRCam Surface Brightness Profile}
\label{appendix:profile}

\begin{figure*}
    \centering
    \includegraphics[width=1\linewidth]{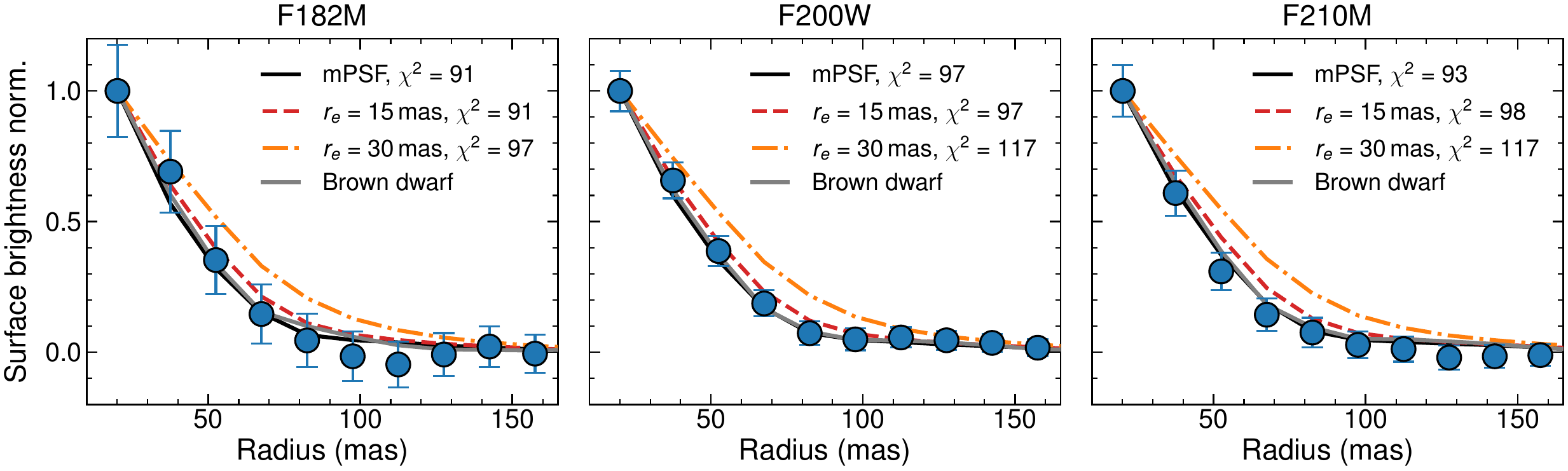}
    \caption{Surface brightness profile of JADES-GS-z14-1 in the NIRCam F182M, F200W, and F210M bands, compared with the model PSF in the mosaic images (mPSF), S\'ersic models with effective radii $r_e=$ 15\,mas and 30\,mas, and a brown dwarf. The mPSF is constructed with {\tt STPSF} using the spectrum of JADES-GS-z14-1. S\'ersic models are generated using the {\tt GALFIT} package assuming a S\'ersic index $n=1$ and zero ellipticity, and are convolved with the mPSF. $\chi^2$ values are computed by fitting the fluxes while keeping the profiles fixed within $0.4''\times0.4''$ cutouts. The brown dwarf is not spatially resolved and has brightness comparable to JADES-GS-z14-1 in these bands.}
    \label{fig:nircam_profile}
\end{figure*}

To illustrate the compactness of JADES-GS-z14-1, we compare its surface brightness profile with the PSF, PSF-convolved S\'ersic models with half-light radii of 15\,mas and 30\,mas, and a brown dwarf, as shown in Figure~\ref{fig:nircam_profile}. We focus on the shortest-wavelength NIRCam bands with detected emission (F182M, F200W, and F210M), where the PSFs are sharpest.

The surface brightness profiles are measured in the mosaic images using concentric annuli with increasing inner radii and a fixed width of 15\,mas (0.5 pixel), normalized to the flux in the central circular aperture of radius 30\,mas (1 pixel). The mosaic PSF model is constructed by generating PSFs for individual exposures using the {\tt STPSF} package, then stacking them onto the mosaic grid while accounting for differences in position angle. When generating the PSFs, we account for the observed spectrum of JADES-GS-z14-1, because these filters partially cover the Ly$\alpha$ break and the flux arises mainly from the longer-wavelength side of each bandpass, which makes the PSF appear slightly wider due to the diffraction law. We use the {\tt GALFIT} software \citep{Peng2002GALFIT, Peng2010GALFIT} to create PSF-convolved S\'ersic models centered on the position of JADES-GS-z14-1. We set the S\'ersic index to $n=1$, as the {\tt ForcePho} posterior for JADES-GS-z14-1 peaks near this value, and galaxies at similar redshifts are generally consistent with $n=1$ \citep[e.g.,][]{Tacchella2023GNz11, Carniani_2024}. We set the ellipticity to zero for simplicity. As Figure~\ref{fig:nircam_profile} shows, in all three bands, the surface brightness profiles of JADES-GS-z14-1 strongly disfavor  models with $r_e = 30$\,mas, but do not clearly distinguish between the point-source model and the model with $r_e = 15$\,mas, except for a slight preference for the point-source model in the F210M band. This is also reflected in the $\chi^2$ metric, obtained by fitting the fluxes while keeping the profiles fixed. Therefore, the surface brightness profile is consistent with a size $\lesssim15$\,mas. We do not present comparisons with smaller radii, as profiles with $r_e < 15$\,mas (0.5 pixel) are not distinguishable in the mosaic images. Attempts at free-parameter {\tt GALFIT} fitting did not converge due to the extremely small size. 

We further validate our PSF model by comparing the surface brightness profiles with those of a brown dwarf (JADES-GS-BD-7; \citealt{Hainline2024ApJ}). This brown dwarf is detected in all three bands with fluxes around 30\,nJy and a red spectrum near 1.8$\,\mu$m, as inferred from the SED fitting \citep{Hainline2024ApJ}, which provides an excellent benchmark for the PSF under faint-source conditions. Its profile in Figure~\ref{fig:nircam_profile} confirms the accuracy of our PSF modeling and further supports the conclusion that JADES-GS-z14-1 is consistent with a point source.

We therefore conclude that the half-light radius of JADES-GS-z14-1 is $\lesssim 15$\,mas (50\,pc). This limit is less stringent than the {\tt ForcePho} constraint of $<10$\,mas (33\,pc) because our test is performed on the mosaic images, whereas {\tt ForcePho} operates on individual exposures. Fitting individual exposures with dithering allows finer sampling and thus provides tighter constraints on the detailed two-dimensional profile. The 15\,mas (50\,pc) upper limit is more conservative but remains critical, as the surface brightness profile offers a direct visual check and rules out the possibility that the compact size is merely the result of extended flux being lost in the noise.
}

\section{Analytical Flux Solution in model-fitting photometry} \label{appendix}

We present the formalism for deriving the optimal flux and its uncertainty for model-fitting photometry, measured directly from individual exposure images. Let the subscript $i$ denote the $i$-th exposure, and each exposure has $n_i$ pixels.  The total chi-squared is given by

\begin{equation} \chi^2 = \sum_i \left(\mathbf{D}_i - \mathbf{A}\mathbf{M}_i - \mathbf{B}_i\mathbf{L}_i\right)\mathbf{C}_i^{-1} \left(\mathbf{D}_i - \mathbf{A}\mathbf{M}_i - \mathbf{B}_i\mathbf{L}_i \right)^\top,
\label{eq:chi2}
\end{equation}
where $\mathbf{D}_i \in \mathbb{R}^{n_i} $ is the vectorized image data for the $i$-th exposure, consisting of $n_i$ pixel values. The pixel covariance matrix $\mathbf{C}_i \in \mathbb{R}^{n_i \times n_i}$ is assumed to be diagonal in exposure images, with each diagonal entry representing the variance from the sigma map. The matrix $\mathbf{M}_i \in \mathbb{R}^{k \times n_i} $ represents the projected models for a total of $k$ galaxies in the $i$-th exposure image, and the flux vector $\mathbf{A} \in \mathbb{R}^k$ is the amplitude of each source, assumed constant across exposures. The elements of $\mathbf{A}$ represent the total fluxes of the sources, provided that each source model is normalized to unit integral over all pixels. The background in each exposure is modeled using a linear combination of basis functions. Specifically, $\mathbf{L}_i\in \mathbb{R}^{3\times n_i}$ contains three background components: a constant and linear gradients along the x and y axes of the images. The vector $\mathbf{B}_i \in \mathbb{R}^{3}$ contains the background coefficients for the $i$-th exposure. The summation symbol denotes a sum over exposures, while summation over pixels is implicitly handled by the matrix multiplications within each term.

At the maximum likelihood estimate, the derivatives of $ \chi^2 $ with respect to variable $\mathbf{A}$ and $\mathbf{B}_i$ vanish, 

\begin{equation}
    \frac{\partial \chi^2}{\partial \mathbf{B}_i^\top} = -2\left(\mathbf{D}_i - \mathbf{A} \mathbf{M}_i - \mathbf{B}_i \mathbf{L}_i \right)\mathbf{C}_i^{-1} \mathbf{L}_i^\top   = 0
    \label{eq:parL}
\end{equation}

Solving Equation~\ref{eq:parL} gives the optimal background coefficients as

\begin{equation}
    \mathbf{B}_i = \left(\mathbf{D}_i -\mathbf{A} \mathbf{M}_i \right) \mathbf{C}_i^{-1}\mathbf{L}_i^\top
    \left(\mathbf{L}_i \mathbf{C}_i^{-1}\mathbf{L}_i^\top \right)^{-1}
\end{equation}

Substituting this expression back into Equation~\ref{eq:chi2}, the chi-squared becomes

\begin{equation}
    \chi^2 = \sum_i \left(\mathbf{D}_i - \mathbf{A}\mathbf{M}_i\right)\mathbf{\tilde{C}}_i^{-1} \left(\mathbf{D}_i - \mathbf{A}\mathbf{M}_i\right)^\top. 
\end{equation}
where the modified inverse covariance matrix, after marginalizing over the background, is given by

\begin{equation}
    \mathbf{\tilde{C}}^{-1}_i = \mathbf{C}_i^{-1} - \mathbf{C}_i^{-1} \mathbf{L}_i^\top \left( \mathbf{L}_i \mathbf{C}_i^{-1} \mathbf{L}_i^\top \right)^{-1} \mathbf{L}_i \mathbf{C}_i^{-1}.
\end{equation}

This formulation reduces the problem to a generalized least-squares estimate. The analytical solution for the flux vector $\mathbf{A}$ is 

\begin{equation}
    \mathbf{{A}} = \left(  \sum_i \mathbf{M}_i\tilde{\mathbf{C}}_i^{-1}\mathbf{M}^\top_i\right)^{-1}\left( \sum_i   \mathbf{M}_i\tilde{\mathbf{C}}_i^{-1}\mathbf{D}_i^\top\right),
    \label{eq:A}
\end{equation}
and the covariance between the fluxes of the $m$-th and $n$-th galaxies is

\begin{equation}
    \mathrm{Cov}({A}_m, {A}_n) = \left(  \sum_i \mathbf{M}_i\tilde{\mathbf{C}}_i^{-1}\mathbf{M}_i^\top\right)^{-1}_{mn}.
    \label{eq:cov}
\end{equation}

This formalism provides a robust and computationally efficient solution for flux estimation in model-fitting photometry. By operating directly on individual exposure images and analytically marginalizing over spatially varying background components, the method accounts for both pixel-level noise and inter-exposure backgrounds. The resulting flux estimates and their covariances are derived through generalized least squares, without relying on iterative numerical optimization. 

\bibliography{reference}{}
\bibliographystyle{aasjournalv7}

\end{document}